\documentclass[aps,prl,twocolumn,superscriptaddress,groupedaddress,amsmath,amssymb,nofootinbib]{revtex4-2}
\usepackage{graphicx}%
\usepackage{bm}        %
\usepackage{amssymb}
\usepackage{booktabs}  %
\usepackage{makecell}  %
\usepackage{tikz}      %
\usetikzlibrary{arrows.meta,positioning}
\usepackage[usenames,dvipsnames]{xcolor}
\usepackage{color}
\usepackage{url}
\RequirePackage[hidelinks]{hyperref}

\def\beq{\begin{equation}}
\def\eeq{\end{equation}}

\newcommand{\Ptm}{P_{\mathrm{tm}}}

\newcommand{\Tpass}{T_{\mathrm{passive}}}
\newcommand{\Tact}{T_{\mathrm{act}}}
\newcommand{\sphase}{\sigma_{\mathrm{phase}}}
\newcommand{\stone}{\sigma_{\mathrm{tone}}}
\newcommand{\gpulse}{g_{\mathrm{pulse}}}
\newcommand{\Tcyc}{T_{\mathrm{cyc}}}
\newcommand{\Ton}{T_{\mathrm{on}}}
\newcommand{\Toff}{T_{\mathrm{off}}}

\begin{document}

\title{An anatomically grounded human-scale dynamic model of contraction-driven lymphatic pumping}

\author{Chia-Feng Chang}
\author{Ayemon Baraka}
\author{Egor Sidorov}
\author{Sabee Molloi}
\email{symolloi@uci.edu}
\affiliation{Department of Radiological Sciences, University of California, Irvine, CA, USA}
\date{\today}

\begin{abstract}
\textbf{Background.} Lymph is driven against an adverse pressure gradient by
the intrinsic contraction of valved muscular segments called lymphangions, and how
these pumps act together across the whole body bears on lymphedema, on immune and
metabolic disease, and on the reading of clinical lymphatic imaging. Existing
models either resolve active lymphangion mechanics in one vessel, a short chain,
or a regional network, or represent the human whole body only as a passive,
reduced-order graph.

\textbf{Methods.} We extend the whole-body passive lymphatic network of the
companion static study \cite{Chang2026Static} to the active, contraction-driven
regime on the same anatomical graph. On every vessel we couple intrinsic
lymphangion contraction, finite wall relaxation, dynamic secondary valves, the
swept-volume source of each shortening segment, and a respiration-gated venous
ostial outlet, advance the unsteady nodal hydraulic system in time, and check the
emergent pressures and flows against published clinical ranges.

\textbf{Results.} Without fitting the dynamic waveform or the contraction
amplitude, the network settles by itself into a bounded, respiration-gated pumping
cycle at the lymphovenous junction, at a clinical operating pressure
($7.5$--$12.3$\,mmHg) and drainage ($35$--$50$\,mm$^3$/s). Two separable signals
appear, a respiration swing of $1$--$2.5$\,mmHg at the junction and a contraction
swing of about $1$\,mmHg inside the duct. Pressure generation is limited more by
its load than by its drive, and no single contraction amplitude fits every vessel.
Raising $P_{\mathrm{subclav}}$ from $3$ to $15$\,mmHg lifts the deep
cervical node from $8.6$ to $21.1$\,mmHg at the weakest drive, a transmission of
$1.04$ that falls to $0.78$ at the strongest, so contraction is what buffers
cervical drainage against venous pressure. These are human cervical pressures
where cervical-scale models are murine.

\textbf{Conclusions.} To our knowledge this is the first model to place
active lymphangion pumping on an anatomically derived human whole-body lymphatic
graph. It turns the pump law of a single lymphangion into a body-level hydraulic
solver, so lymphedema and lymphatic imaging can be read in terms of vessel
mechanics. The same solver computes the cervical outlet pressure that
cerebrospinal-fluid drainage models prescribe.
\end{abstract}

\maketitle

\section{Introduction}
\label{sec:intro}

Lymph is transported against an adverse pressure gradient principally by the
intrinsic, valve-gated contraction of muscular vessel segments called
lymphangions \cite{Zawieja2009,Davis2025,Davis2009,Davis2011}. Each
lymphangion behaves as a miniature pressure-driven pump. Circumferential smooth
muscle generates a phasic active tension, secondary bicuspid valves rectify the
resulting flow, and chains of such units in series produce net forward transport
\cite{Scallan2016,Breslin2019}.

How these pumps act together at the scale of the whole body bears on both
disease and imaging. Pump failure underlies lymphedema and is implicated in
immune and metabolic disorders \cite{Scallan2016,MooreBertram2018}, and the
clinical readouts are network-level, since lymphoscintigraphy and near-infrared
imaging track transport across many vessels at once. The vessels are at the same
time narrow, long, and hard to access, so a large part of the network resists
direct measurement.

Quantitative models have been central to interpreting these experiments,
and they are for the most part reduced-order. For example, Reddy et al., \cite{Reddy1977} built the
first whole-body one-dimensional model of lymphatic flow.  Quick et al., \cite{Quick2007,Quick2008} represented the spontaneously contracting lymphangion as a lumped
circuit element, and Bertram, Jamalian, and co-workers \cite{Bertram2011,Bertram2014,Bertram2016,Bertram2018,Jamalian2013,jamalian2016network,Contarino2017,Razavi2017}
developed chains of a few lymphangions with compliant walls and hysteretic valves.
These lumped models fixed the single-vessel pump physics in quantitative detail
\cite{MargarisBlack2012}.

A second line of work resolves the vessel wall and the valve leaflets in two
or three dimensions through fluid-structure interaction
\cite{Rahbar2011,Baish2016,Li2019,Elich2021,Ballard2018,Wolf2021}. These models
capture valve mechanics, contraction timing, and valve-level backflow in detail.
Their computational cost, however, confines them to a single lymphangion or a
short chain.

Both lines share a limitation of scale. Their parameters come mostly from
the thoracic duct and mesenteric vessels, and neither resolves how body-wide
topology, the hierarchical distribution of vessel calibre, and a realistic venous
boundary jointly shape the pulsatile pressure and flow fields. Network-scale
models have begun to supply the missing geometry, through anatomical lymphatic
graphs \cite{Savinkov2020,tretyakova2017developing}, meshes of initial lymphatics
\cite{ikhimwin2020computational}, and recent lumped-parameter models of regional
networks such as the cervical lymphatics \cite{KimTithof2024}. Closely
related network models describe cerebrospinal-fluid clearance through the
glymphatic system, representing the brain-wide perivascular pathway as a
hydraulic-resistance network \cite{Tithof2022,Bohr2022} and coupling it to
blood-brain-barrier transport \cite{yousofvand2025numerical}. These remain
passive, regional, or non-human. A whole-body, anatomically explicit model of
active human lymphatic pumping is still missing. 

In the companion study \cite{Chang2026Static}, we built the passive substrate for
this question. On a whole-body anatomical graph closed by a dual-derivation Murray
radius law, we solved the steady passive network with an experimentally informed
wall law, hysteretic valves, transmural leak, lymph-node retention, and a
three-element venous outlet, with the active muscle tension switched off. That
baseline fixes the hydraulic skeleton but is, by construction, silent on the
defining feature of lymph transport, intrinsic contraction.

Here we activate the contractile machinery on that same network. The new physics
add five pieces. (i)~A phasic active muscle tension assembled from a
distension-gated length-tension window, a per-region contraction waveform, and a
single global phasic-tension scale. (ii)~A first-order wall-relaxation dynamics
that lets the luminal radius track its mechanical equilibrium with a finite time
constant. (iii)~Dynamic valve gating driven by the instantaneous
transvalvular pressure. (iv)~A contraction-volume source that injects the swept
volume of each shortening segment into the nodal mass balance. (v)~An ostial valve
at the venous junction, distinct from the intraluminal valves and physiologically
respiration-gated, that gates the outlet between an open-state ostial resistance
on inspiration and a closed no-flux condition on expiration over the breath cycle.
The coupled system is advanced in time to a quasi-steady pumping cycle.

We make two deliberate scope restrictions. First, the phasic-tension amplitude
$\sphase$ (together with the valve steepness and the contraction phase lag)
is held at a single \emph{globally constant} value instead of fitted per vessel.
Their per-edge calibration ($\sphase$ from a wall-compression target, the valve
steepness from a retroflow target) is the subject of a separate study. Second, contrast-tracer transport, the
time-activity readout most directly comparable to lymphoscintigraphy and
ICG/NIRF imaging, is likewise reserved for a companion paper. The present
contribution is the dynamic transport model itself: its constitutive
construction, its numerical solution on a human-scale network, and the emergent
pumping behaviour it predicts.

\section{Methods}
\label{sec:methods}

\subsection{Static substrate and notation}
\label{sec:substrate}

The dynamic model is built on the converged passive network of the companion
study \cite{Chang2026Static}, whose construction we summarize only to fix
notation. An anatomically derived human lymphatic graph
\cite{Savinkov2020,tretyakova2017developing} is expanded to a three-level mesh:
each anatomical edge is divided into $N_{\mathrm{lym}}$ series lymphangions
(a uniform inter-valve spacing, i.e.~the lymphangion length,
$s_{\mathrm{valve}}=3$\,mm applied across all regions), and each lymphangion into
$N_g=5$ wall sub-segments, giving $\sim$89{,}060 sub-edges.

The human inter-valve spacing distribution is not established, and the sparse
data are surveyed in the companion study \cite{Chang2026Static}. This
spacing
sets the number of series pumping units on an anatomical edge of length $\ell$,
$N_{\mathrm{lym}}=\max\!\big(1,\operatorname{round}(\ell/s_{\mathrm{valve}})\big)$,
hence the valve density and the swept volume per contraction, so an
order-of-magnitude change in $s_{\mathrm{valve}}$ would rescale the pump. Lacking
organ-resolved human data, we use a uniform $3$\,mm here and defer an organ-wise
survey of $s_{\mathrm{valve}}$ to future work.

Vessel radii are set by the dual-derivation Murray closure and the steady passive
solve provides, on every sub-edge $e$, an operating-point radius $r_e$, transmural
pressure $\Ptm$, and flow. Quasi-steady laminar flow $Q_e$ on each edge obeys Hagen--Poiseuille,
\begin{equation}
  Q_e = g_e\,(p_i-p_j), \qquad
  g_e = \frac{\pi r_e^{4}}{8\,\eta\,L_e\,(1+\phi_e)},
  \label{eq:poiseuille}
\end{equation}
with $\eta=1.8\times10^{-3}$\,Pa$\cdot$s and $\phi_e$ the dimensionless valve
resistance factor reflects whether the valve is open or close (Sec.~\ref{sec:valve_dyn}). Here $g_e$ is the edge
conductance, $L_e$ the vessel length of edge $e$, and $p_i,p_j$ the pressures at
its two endpoints $i,j$.

The passive wall law is a parameterized exponential relation on the vessel tension \cite{Athanasiou2017}
\begin{equation}
  \Tpass(r) = k_1\!\left[\exp\!\big(k_2(r/r_c-1)\big)-1\right],
  \label{eq:wall}
\end{equation}
where $r$ is the actual luminal radius of the wall sub-segment, $k_1$ is the wall tension scale, $k_2$ the dimensionless stiffness exponent, $r_c$ is the per-edge reference radius at which the passive tension
vanishes, calibrated therefore $\Tpass$ reproduces the static operating point. We adopt the parameters of $k_1$, $k_2$, and $r_c$ as in \cite{Chang2026Static}. The transmural
pressure carried by the wall is the thin-wall Laplace relation
$P = P_{\mathrm{anchor}}+\Tpass/r$, truncated at zero in the
tensionless regime ($\max(P_{\mathrm{anchor}}+\Tpass/r,0)$), since a
tensionless, fully relaxed wall is mechanically slack and exerts no outward
pressure on the lumen. The $P_{\mathrm{anchor}}$ helps for numerical stability and it is ignorable in value. Splitting each vessel into many
sub-edges also means its macroscopic response is not a single Poiseuille law.
Each sub-edge is locally linear, but its conductance depends on the local radius
through $g_e\propto r_e^{4}$ and the wall sets that radius from the local
transmural pressure, so the chain of sub-edges makes the whole-vessel flow a
nonlinear function of the end-to-end pressure drop, see figure.~1 in \cite{Chang2026Static}.

\subsection{Active muscle tension}
\label{sec:active}

The contractile machinery enters as an additional active circumferential
tension $\Tact$ that adds to the passive wall tension in the radius equilibrium.
Following the decomposition in study \cite{Caulk2015,Razavi2020,Davis2025}, $\Tact$ is the product of a wall
geometric factor, a length--tension gate, and a temporal contraction waveform:
\begin{align}
\label{eq:Tact}
  \Tact^{(e)}(t) &= h(R_e)\;\Phi(\lambda_{\theta,e})\;f_{\mathrm{SNP}} \\ \notag & \times 
    \left[\,\stone^{(e)} + \sphase \; \gpulse^{(e)}(t)\,\right]
    g_{\mathrm{spatial}}(z),
\end{align}
where $t$ is simulation time,
$h(R_e)$ is the calibre-dependent wall thickness, $\Phi(\lambda_{\theta,e})$ the
length-tension window of the circumferential wall stretch $\lambda_{\theta,e}$,
$f_{\mathrm{SNP}}$ the nitric-oxide-mediated relaxation factor, $\stone^{(e)}$ and
$\sphase$ the tonic and phasic muscle-stress scales, $g_{\mathrm{pulse}}^{(e)}(t)$
the per-edge contraction waveform, and $g_{\mathrm{spatial}}(z)$ the axial force
profile along the lymphangion coordinate $z$, each specified below.

\paragraph{Wall thickness $h(R)$.}
The active stress acts over the muscular wall, whose thickness scales with vessel
calibre. From representative collecting-lymphatic wall-thickness measurements
spanning initial lymphatics to the thoracic duct (Table~\ref{tab:hwall}) we adopt
a linear calibre dependence,
\begin{equation}
  h(R) = (37.0\pm7.4) + (8.3\pm7.9)\,R\quad[\mu\mathrm{m}],
  \label{eq:hwall}
\end{equation}
with $R$ the hydraulic radius in mm. The linear trend is weak
($R$-square $\approx0.15$), reflecting the heterogeneous multi-source compilation. At a representative $R=1$\,mm this gives
$h\approx45\,\mu$m. The compiled data are heterogeneous in species and protocol
(cadaveric histology, unloaded geometry, controlled pressure--diameter testing,
and in vivo remodelling) and are best read as morphology-dependent reference
values, which is reflected in the
large relative uncertainty on the slope (the individual sources are listed
in Table~\ref{tab:hwall}). Because $h$ enters Eq.~(\ref{eq:Tact}) linearly, the
active tension inherits this calibre scaling of the wall.

\begin{table*}[t]
  \centering
  \caption{Representative lymphatic-vessel wall thicknesses compiled for the
  $h(R)$ fit of Eq.~(\ref{eq:hwall}). $D$ is lumen diameter and $R=D/2$ the lumen
  radius. The cited studies use heterogeneous protocols (anatomical description,
  fixed histology, unloaded geometry, controlled pressure--diameter testing, in
  vivo remodelling), so the values are morphology-dependent reference data rather
  than a single pressure-matched set. The initial-capillary row is structural
  (no collecting-vessel muscular wall) and is excluded from the fit.}
  \label{tab:hwall}
  \begin{tabular}{lllll}
    \toprule
    Vessel type & Size & Wall thickness $h$ & Condition & Ref. \\
    \midrule
    Initial / terminal lymphatic capillary & $D\sim10$--$60\,\mu$m & endothelial monolayer & anatomical & \cite{Scallan2010} \\
    Human meningeal lymphatic-like & $D\approx27\,\mu$m & $5.3\pm2.8$; $7.2\pm5.0\,\mu$m & fixed histology & \cite{VeraQuesada2023} \\
    Human lower-limb collecting & $D\approx0.3$\,mm & $51.3\pm12.6\,\mu$m & ex vivo, unloaded & \cite{Gianesini2023} \\
    Bovine mesenteric collecting & $D\approx2$\,mm & $\sim$30\,$\mu$m & $0$--$30$\,cmH$_2$O & \cite{Arkill2010} \\
    Rat thoracic duct & $R_{\mathrm{unloaded}}\approx240\,\mu$m & $32.2\pm1.6\,\mu$m (unloaded) & $P_{\mathrm{tm}}=3$--$9$\,cmH$_2$O & \cite{Caulk2015} \\
    Canine thoracic duct & -- & $h/R_o\approx3.5\%$ & $2$--$6$\,cmH$_2$O & \cite{Deng1999} \\
    Swine thoracic duct (control) & $D=3.35\pm0.37$\,mm & $60\pm10\,\mu$m & in vivo & \cite{Lu2022} \\
    Swine thoracic duct (elevated CVP) & $D=4.32\pm0.57$\,mm & $260\pm20\,\mu$m & in vivo remodelling & \cite{Lu2022} \\
    \bottomrule
  \end{tabular}
\end{table*}

\paragraph{Length--tension window $\Phi(\lambda_\theta)$.}
Active tension is gated by the local circumferential stretch through a length--tension window \cite{Zhang2007,Davis2009,Caulk2015,Razavi2020},
\begin{equation}
  \Phi(\lambda_\theta) = \lambda_\theta\,
    \max\!\left[0,\; 1 - \left(\frac{\lambda_M-\lambda_\theta}{\lambda_M-\lambda_0}\right)^{\!2}\right],
  \label{eq:phi}
\end{equation}
which is vanishing/weakening outside the active range
$[\lambda_0,\lambda_M]$. The stretch variable is referenced to the wall-law
reference radius,
\begin{equation}
  \lambda_\theta = \lambda_0 \frac{r}{r_c}.
  \label{eq:lambdatheta}
\end{equation}
where we adopt $\lambda_0=0.25$ and $\lambda_M=1.10$.
The lower bound $\lambda_0$ is reached at the wall-law reference radius
$r_c$ of the companion passive study \cite{Chang2026Static}, where the passive
tension also vanishes, so both tensions share one fully relaxed reference and no
active tension develops under further compression $r\le r_c$. The upper bound
$\lambda_M$ places peak activation at $r=(\lambda_M/\lambda_0)\,r_c=4.4\,r_c$,
above the physiological range, so within that range $\Phi$ decreases
monotonically toward $r_c$ and brakes over-compression. %

\paragraph{Tonic stress.}
The bracket in Eq.~(\ref{eq:Tact}) splits the muscle stress into a tonic and a
phasic part. The tonic term is
\begin{equation}
  \stone = a_{\mathrm{tone}}\,\max(P_{\mathrm{tm}},0).
  \label{eq:sigmatone}
\end{equation}
In the present model the tonic coefficient is set to zero
($a_{\mathrm{tone}}=0$), so the active tension is carried entirely by the phasic term. This tonic term is written as $\stone\propto P_{\mathrm{tm}}$, so it
scales with the same operating pressure that the passive wall law already carries
through $r_c$ and is degenerate with a shift of $r_c$. A nonzero
$a_{\mathrm{tone}}$ would therefore double-count the pressure-dependent operating
tension, whereas setting it to zero lets $r_c$ carry the offset and leaves the
phasic term as the only genuinely dynamic contribution. This matches the
physiology, where the spontaneous phasic contraction dominates the lymphatic
active response \cite{Telinius2010}.

\paragraph{Phasic stress $\sphase$.}
The phasic stress scale $\sphase$ is the amplitude of the contraction-driven
tension~\cite{Telinius2010,Zhang2013}. We treat $\sphase$ as a single globally constant parameter,
identical for every vessel within a run, and report results across a sweep of its
value. The literature species-scaling anchor is
\begin{equation}
  \sphase^{\mathrm{lit}} = \sphase^{\mathrm{mouse}}\,\alpha_p^{\mathrm{mouse}}
          = 1650~\mathrm{Pa}\times 3.3 \approx 5445~\mathrm{Pa},
  \label{eq:sigmaphase}
\end{equation}
with $\sphase^{\mathrm{mouse}}=1650$\,Pa a representative lymphatic phasic
active-tension scale \cite{Razavi2020,Caulk2015} and $\alpha_p^{\mathrm{mouse}}\simeq 3.3$ the mouse$\to$human pressure-scale
ratio \cite{Chang2026Static} (Sec.~\ref{sec:species}). The \emph{per-vessel} calibration of $\sphase$
against regional pressure-amplitude data is out of scope here and is treated in a
companion future study. A single global value per run isolates the dynamic-transport
behaviour from the freedom of a spatially varying contraction amplitude.

\paragraph{NO-mediated relaxation $f_{\mathrm{SNP}}$.}
The factor $f_{\mathrm{SNP}}$ models nitric-oxide--mediated vasorelaxation. It is
held at unity ($f_{\mathrm{SNP}}=1$) in the present runs and retained for future
pharmacological perturbation studies.

\paragraph{Axial force distribution $g_{\mathrm{spatial}}(z)$.} 
Within a lymphangion the active stress is distributed along the fractional
axial coordinate $z\in[0,1]$. We consider two normalized profiles and apply them with finite element algorithm (as in the figure~1 in \cite{Chang2026Static}): a \emph{uniform} distribution
($g_{\mathrm{spatial}}=1$) and the production \emph{sinusoidal} profile
$g_{\mathrm{spatial}}(z) = 2\sin^2(\pi z)$, the later profile vanishes at the
two valve ends (the muscle does not contract against the rigid valve seats) and
peaks at mid-lymphangion. Both integrate to unity over the lymphangion,
$\langle g_{\mathrm{spatial}}\rangle_z = 1$, so the sub-segment-averaged muscle
stress is fixed to $\sphase$ independent of the profile. The choice between
the uniform and $\sin^2$ profiles have only a minor effect on the
pumping observables in our simulation test. We adopt the uniform profile.

\paragraph{Active and passive lymphangions.}
A lymphangion is held passive ($\Tact=0$) when it cannot rectify the flow it
generates, which happens in three cases. (i) A short edge $L_e \lesssim 3$~mm
carries only one valve, so backflow leaks through its unvalved end. (ii) An even
shorter edge $L_e \ll 3$~mm carries no valve at all. (iii) The segment abutting a
graph junction carries no valve.

In the model, valves sit only at the $N_{\mathrm{lym}}-1$ interior boundaries
of each edge, so the two end lymphangions abut a valveless graph vertex and are set
passive, while only the $N_{\mathrm{lym}}-2$ interior lymphangions contract (an edge
with $N_{\mathrm{lym}}\le2$ then pumps nothing). Activating the ends as well would
raise the thoracic-duct contraction amplitude by about 25\%, but that gain is
spurious because an unrectified end cannot sustain net forward flow.

\subsection{Contraction cycle and pumping waveform}
\label{sec:cycle}

\begin{figure}[t]
  \centering
  \includegraphics[width=1\linewidth]{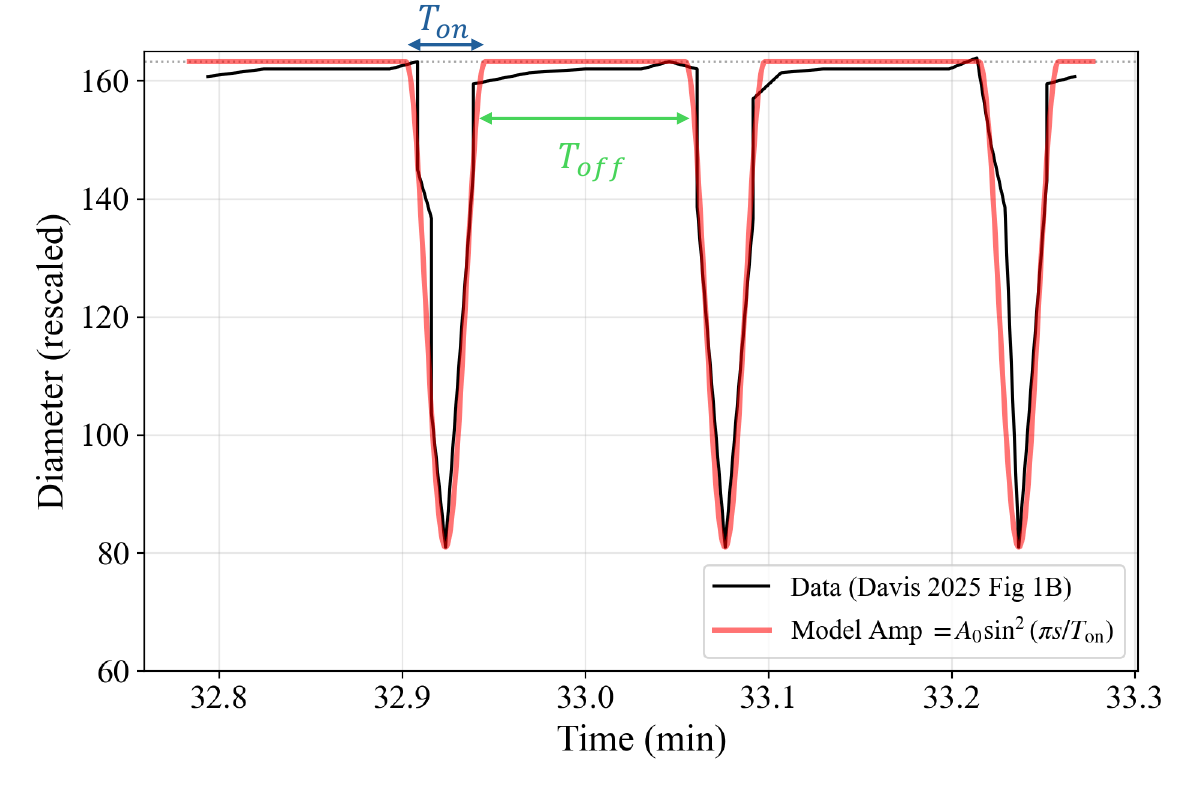}
  \caption{Spontaneous contraction waveform of a mouse lymphatic vessel. The
  trace is the measured luminal diameter over three consecutive contraction
  cycles, from Davis et~al.\ \cite{Davis2025} Fig.~1B for a baseline,
  flow-unresponsive vessel. We fit the raised-sine phasic gate $g_{\mathrm{pulse}}$
  of Eq.~(\ref{eq:gpulse}). }
  \label{fig: Davis1B3p}
\end{figure}

\paragraph{Waveform.}
Each lymphangion contracts on a cycle of period $\Tcyc = \Ton + \Toff$,
comprising an active systole $\Ton$ and a refractory diastole $\Toff$. The
phasic gate is a raised-sine pulse \cite{Davis2025},
\begin{equation}
  \gpulse(t) =
  \begin{cases}
    \sin^2\!\left(\pi \dfrac{s_{\mathrm{pulse}}(t)}{\Ton}+\theta_e\right),
      & 0 \le s_{\mathrm{pulse}}(t) < \Ton,\\[4pt]
    0, & \Ton \le s_{\mathrm{pulse}}(t) < \Tcyc,
  \end{cases}
  \label{eq:gpulse}
\end{equation}
where $\theta_e$ is the initial phase of edge $e$ (Sec.~\ref{sec:phase}), and $s_{\mathrm{pulse}}(t) = t - \lfloor t/\Tcyc \rfloor\,\Tcyc$
is the time elapsed since the start of the current cycle.
We adopt a raised-sine pulse because it is the simplest smooth, non-negative
shape that rises from zero, peaks once, and returns to zero within the active
window, which represents a single systolic contraction followed by a silent
diastole. We further assume that every contraction has the same amplitude, so the
beat-to-beat variability enters only through the timescales $\Ton$ and $\Toff$ and
not through the peak height. We fix these two timescales by fitting
Eq.~(\ref{eq:gpulse}) to the baseline spontaneous diameter trace of a
flow-unresponsive mouse vessel (Davis et al.\ \cite{Davis2025} Fig.~1B, Fig.~\ref{fig: Davis1B3p}). A flow-unresponsive trace isolates the intrinsic
rhythm, because imposed flow suppresses the contraction frequency and amplitude of
isolated lymphatics \cite{Gashev2002}, and the sharp raised-sine systole is
consistent with the fast shortening kinetics of lymphatic muscle \cite{Zhang2013}.
We report the resulting mouse $\Ton$ and $\Toff$
in Table~\ref{tab:tcyc}.

\paragraph{Per-region frequency and duty.} 
The contraction frequency is region-specific, taken from the reported
per-territory pumping rates (Table~\ref{tab:tcyc}). The duty ratio
$\Ton/\Tcyc$ is not resolved per region. We adopt the single mouse value
$\Ton/\Toff = 2.61/6.28$, i.e.\ $\Ton/\Tcyc\approx0.293$ \cite{Davis2025}, uniformly across
regions, and split each region's measured $\Tcyc$
accordingly. This combines region-specific frequency data with a universal duty
cycle, avoiding the claim of region-wise duty measurements that are not observed yet.

\begin{table}[t]
  \centering
  \caption{Per-region contraction frequency and the derived cycle, systole, and
  diastole times. A full macro-scale observation set is compiled in Table~\ref{tab:lymph_pumping_frequency_updated}. The duty split uses the uniform
  Davis-mouse ratio $\Ton/\Tcyc=0.293$.}
  \label{tab:tcyc}
  \begin{tabular}{lrrrrl}
    \toprule
    Region & freq & $\Tcyc$ & $\Ton$ & $\Toff$ & ref. \\
           & (cyc/min) & (s) & (s) & (s) & \\
    \midrule
    Thoracic duct & $5.0\pm1.0$  & 12.00  & 3.52  & 8.48  & \cite{Kelly2022} \\
    GI / mesenteric & $3.4\pm1.7$ & 17.65 & 5.18  & 12.47 & {\cite{Telinius2017}} \\
    Upper limb    & $0.9\pm0.4$  & 66.67  & 19.57 & 47.10 & {\cite{Kelly2020}} \\
    Head / neck   & $0.65\pm0.21$& 92.31  & 27.10 & 65.21 & {\cite{Kelly2020,Mohanakumar2020,Mohanakumar2021,HolmWeber2022}} \\
    Lower limb    & $0.40\pm0.12$& 150.00 & 44.04 & 105.96& {\cite{Mohanakumar2020,Mohanakumar2021,HolmWeber2022}} \\
    \bottomrule
  \end{tabular}
\end{table}

\paragraph{Pressure--frequency coupling.}
The cycle frequency rises with transmural pressure following the saturation law \cite{Zawieja2018},
\begin{equation}
  f_{\mathrm{cyc}}(\Ptm) = \frac{1}{\Tcyc}
    = f_{\mathrm{terminal}}\left[1-\exp(-\mu_p\max(\Ptm,0))\right].
  \label{eq:omega}
\end{equation}
We fit this saturation law to mouse pressure-frequency data \cite{Zawieja2018},
which fixes the terminal frequency $f_{\mathrm{terminal}} \sim 18$ cycles per
minute and the pressure sensitivity $\mu_p^{\mathrm{mouse}} \sim 2.04$\,cmH$_2$O$^{-1}$ with $R^2=0.93$. However, we tried to switch this coupling on in the dynamic runs, but $\mu_p$ is not yet
well determined for human and it sits inside the exponential. It makes the pump equations
numerically stiff and can destabilize the coupled solve. We therefore defer the
effect. The baseline runs hold each region at its terminal, pressure-saturated
frequency $f_{\mathrm{cyc}}=f_{\mathrm{terminal}}$, and we leave the
pressure-tracking coupling for future work once $\mu_p$ is measured more precisely.

\subsection{Contraction phase organization}
\label{sec:phase}

In this paper the lymphangion phases are initialized in-phase
($\Delta\phi=0$) and left uncoupled. The framework also admits a peristaltic
initial stagger and a nearest-neighbour Kuramoto coupling \cite{Kuramoto1975},
  \begin{equation}
    \frac{d\theta_i}{dt} = \omega_i
      + K_{\mathrm{couple}}\sum_{j} A_{ij}\sin(\theta_j-\theta_i),
    \label{eq:kuramoto}
  \end{equation}
where $\theta_i$ is the phase of lymphangion $i$, $K_{\mathrm{couple}}\ge 0$ is a single scalar coupling strength in rad/s, and $A_{ij}$ is the degree-normalized lymphangion adjacency.  No measurement fixes those parameters. %
  
Even from an in-phase start the contractions desynchronize: because each
lymphangion runs at its own pressure-set cycle frequency, $\omega_i = 2\pi f_{\mathrm{cyc}}$
(Eq.~\ref{eq:omega}), an initially common phase disperses over time and the
network phase-order parameter $r_{\mathrm{all}}(t)=|\langle e^{i\theta_i}\rangle|$
settles low, i.e.\ no
global synchrony emerges. The pumping state below is therefore one of
quasi-independent contractions.

We keep zero phase lag and $K_{\mathrm{couple}}=0$ as the simplest baseline
for a single thoracic duct. At fixed active tension this synchronous drive
maximizes the coherent volume swing of the duct, so it behaves like an ideal
positive-displacement pump that ejects its full stroke on each systole and refills
it on each diastole. In this reduced single-duct picture that gives the highest
mean flow. This limit is a modeling choice, not a general optimum.

A five-lymphangion murine cervical model \cite{KimTithof2024} instead
prescribes a $90^\circ$ delay between adjacent lymphangions, a value taken from
earlier chain models, and listed there among its own
limitations. Smooth-muscle coverage is what
makes a collecting vessel contract at all, and vessels that lose it dilate and
stop contracting \cite{wang2017smooth}. The zero-lag baseline here and
the $90^\circ$ there are two settings of the same parameter, so the two models
differ in setup and not in result.

The healthy animal sits near the zero-lag end. Cervical lymphatics in
control mice contract as fully entrained waves at a conduction speed of about
$10$\,mm\,s$^{-1}$, with a single dominant frequency near $10$\,min$^{-1}$
\cite{Hussain2023}. Over a $2$\,mm lymphangion \cite{KimTithof2024} that is a
delay of $0.2$\,s, about $12^\circ$ of the $6$\,s cycle. A chain model that
swept the delay directly found the mean flow to change by $5.1\%$ between
orthograde and retrograde propagation once the delay is under $1$\,s
\cite{Venugopal2007}, and the delay estimated here is well inside that range.
The systematic study of the phase lag
across geometries is left to future work.

\subsection{Wall relaxation dynamics}
\label{sec:wallode}

In the static substrate the radius is the instantaneous solution of the Laplace
equilibrium. Dynamically, the luminal radius relaxes toward its
pressure- and tension-dependent target with a finite time constant
$\tau_{\mathrm{wall}}$,
\begin{equation}
  \tau_{\mathrm{wall}}\,\frac{dr}{dt} = r_{\mathrm{target}} - r,
  \label{eq:wallode}
\end{equation}
where the target $r_{\mathrm{target}}$ is the quasi-static
radius the wall would reach if the current transmural pressure $\Ptm$ and active
tension $\Tact$ were held fixed, given by the instantaneous Laplace balance
$r_{\mathrm{target}}\,\Ptm = \Tact + \Tpass(r_{\mathrm{target}})$. The radius $r$
relaxes toward this moving target with time constant $\tau_{\mathrm{wall}}$ and
coincides with it only in the static limit. 

Eq.~(\ref{eq:wallode}) is the overdamped limit of a wall element modeled
as a damped harmonic oscillator, $m\ddot r + b\dot r + k\,(r-r_{\mathrm{target}})=0$,
with wall inertia $m$, viscous damping $b$, and stiffness $k$. A lymphangion
contraction lasts less than a second, so the wall-mass inertia is negligible and the
wall inertial time $\tau_{\mathrm{inert}}=\sqrt{m/k}$ tends to zero. The oscillator then
reduces to the first-order viscous-elastic balance as Eq.~(\ref{eq:wallode}). The time constant $\tau_{\mathrm{wall}} = b/k$ is therefore the ratio of wall viscous
damping to wall stiffness, where the damping collects cross-bridge cycling and tissue
viscoelasticity. 

We integrate Eq.~(\ref{eq:wallode}) with a second-order backward-difference
(BDF2) scheme,
\begin{equation}
  r^{n+1} = \frac{2\,r^{n} - \tfrac{1}{2}\,r^{n-1}
      + \gamma\,r_{\mathrm{target}}^{n+1}}{\tfrac{3}{2}+\gamma},
  \qquad \gamma \equiv \frac{\Delta t}{\tau_{\mathrm{wall}}},
  \label{eq:wallbdf2}
\end{equation}
where $n$ indexes the pump time step and $\gamma$ is the step size in units of
the wall relaxation time. The source $r_{\mathrm{target}}$ is pre-smoothed by a matched
low-pass that damps high-frequency numerical noise, and
the production run uses $\tau_{\mathrm{wall}}=0.333$\,s and $\Delta t=0.1$\,s.
Since $\tau_{\mathrm{wall}}$ has no direct lymphatic measurement, we fix it to the
smooth-muscle force-decay scale $c_2^{-1}\approx0.333$\,s of Contarino and Toro
\cite{Contarino2017}, within the $0.1$ to $1$\,s smooth-muscle range. The limit
$\tau_{\mathrm{wall}}\to0$ recovers the instantaneous Laplace equilibrium assumed by
lumped lymphangion models. At $\gamma=0.3$ the discretization keeps about $97\%$ of the
contraction amplitude with a phase lag of about $13^\circ$ at the thoracic-duct period,
so it does not materially distort the pumping waveform. We will discuss more in the phenomenon later in Sec.~\ref{sec:results}.

\paragraph{Sub-segment axial coupling.}
To keep the discretized wall continuous, we couple the $N_g$ sub-segments of
each lymphangion (Sec.~\ref{sec:substrate}) elastically along the vessel axis, so the
wall behaves as one elastic tube. After each
wall-ODE step the sub-segment radii are smoothed by a discrete Laplacian,
\begin{equation}
  r_k \;\leftarrow\; r_k + \alpha_r\,(r_{k-1} + r_{k+1} - 2 r_k),
  \label{eq:alphar}
\end{equation}
along the sub-segment index $k$ within each lymphangion (Neumann boundary at the
lymphangion ends). We use $\alpha_r=0.2$ throughout this paper, within the
numerically stable range $\alpha_r\in[0,0.5]$. This coupling is not cosmetic. At
$\alpha_r=0$ the sub-segments relax independently, and their out-of-phase local
volume changes act as a spurious per-segment Windkessel that substantially damps the
cyclic pressure pulse once $N_g>1$. Raising $\alpha_r$ to $0.2$ restores the axial
continuity and recovers the pulse, so the result depends strongly on $\alpha_r=0$
versus $\alpha_r>0$.

\paragraph{Picard coupling.}
Each simulation time step couples the active tension, the wall radius, and the passive
wall pressure, and we resolve that coupling to a self-consistent fixed point by
Picard iteration. Evaluating the three in a single pass would leave a one-step lag,
so instead each Picard sweep applies the coupled update map
\begin{equation}
  \stone \to \Tact \to r_{\mathrm{target}} \to \text{wall step}
    \to \langle\gpulse\rangle \to P_{\mathrm{passive}},
  \label{eq:picard}
\end{equation}
feeding the updated $P_{\mathrm{passive}}$ into the next sweep. Repeating the
map drives the coupled state to its fixed point, and we run $N_{\mathrm{Picard}}=5$
sweeps per step.

\paragraph{Fluid inertance.}
Lymph carries mass, so an accelerating fluid column resists the pressure
gradient that drives it, and we must check whether that inertia matters over a
contraction cycle. Keeping it replaces the instantaneous Poiseuille law of
Eq.~\eqref{eq:poiseuille} by an unsteady momentum balance on each
edge~\cite{Womersley1955},
\begin{equation}
  L_h\,\frac{dQ_e}{dt} + \mathcal{R}_e\,Q_e = \Delta p_e,
  \qquad L_h = \frac{\rho\,L_e}{A_e},
  \label{eq:inertance}
\end{equation}
with lymph density $\rho=1060$\,kg\,m$^{-3}$, $A_e=\pi r_e^2$ the
lumen cross-section, and $\mathcal{R}_e=1/g_e$ the edge hydraulic
resistance of Eq.~(\ref{eq:poiseuille}). The new term makes the flow
lag the pressure drop by the inertial time constant
$\tau_e = L_h/\mathcal{R}_e = \rho\,r_e^2/(8\eta)$, which depends on the radius alone and
runs from $10^{-4}$\,s in the peripheral collectors to $2\times10^{-1}$\,s in the
thoracic duct. Every value lies far below the contraction period
$\Tcyc\sim10$--$150$\,s, so the flow tracks the pressure gradient almost
instantaneously and the inertial term stays small at the pumping frequency.

We estimate the size of this correction with the per-edge Womersley number $W_o^{(e)} = r_e\sqrt{2\pi\rho/(\eta\Tcyc)}$ \cite{Womersley1955}, which
sets the length-free ratio
\begin{equation}
  \omega\,\tau_e = \frac{\bigl(W_o^{(e)}\bigr)^2}{8},
  \qquad \omega \equiv \frac{2\pi}{\Tcyc}.
  \label{eq:womersley}
\end{equation}
At thoracic-duct calibre ($r\approx1.5$\,mm) and the production period ($\Tcyc\approx9$--$12$\,s) the duct
sits at $W_o^{(e)}\approx0.83$--$0.96$, hence $\omega\tau_e\approx0.09$--$0.12$,
below $1\%$ in magnitude and $5$--$7^\circ$ in flow-to-gradient phase lag. In vivo particle
tracking gives $W_o^{(e)}\approx0.04$ in the much narrower mouse cervical
lymphatics~\cite{du2024restoration}. The ratio grows as
$r_e^2/\Tcyc$, so the term is not equally small in every vessel. A lymphangion
model of the human thoracic duct gives a larger
$W_o^{(e)}\approx1.4$~\cite{Rahbar2011}, which looks too large to neglect. Our
estimate shows that its effect is still small.

\subsection{Contraction-volume source and valve dynamics}
\label{sec:valve_dyn}

\paragraph{Pumping source.}
As each sub-segment changes radius it sweeps a volume rate
$dV_e/dt = 2\pi r_e L_e\,(dr_e/dt)$, which enters the nodal mass balance as a
source. The wall contraction has no preferred axial direction, so we deliver the swept volume equally\footnote{It may not be equal in reality.}
to the two ends of the sub-segment,
\begin{equation}
  Q_i^{(e)} = Q_j^{(e)} = -\frac{1}{2}\,\frac{dV_e}{dt},
  \label{eq:src}
\end{equation}
where $Q_i$ and $Q_j$ are the flows the wall delivers to the two endpoints
$i$ and $j$ of sub-segment $e$, and the sign of $dV_e/dt$ selects the contraction
or relaxation half-cycle. This
equal split carries no direction of its own, and
over a closed cycle the wall returns to its starting radius, but the valves
supply the direction. They block the backward half of each stroke and pass the
forward half, so all net transport comes from the phase-aware gating described
below.

The radius does not simply follow the active tension, because two things
resist the squeeze. The lymph inside must be driven out through the network
resistance, which raises the transmural pressure, and the passive wall tension
$\Tpass$ pushes back against the same pressure. We therefore need the
pressure-radius relation of the wall, which we carry as a lumped compliance
\begin{equation}
  C_e = \frac{dV_e}{d\Ptm} = 2\pi r_e L_e\,\frac{dr_e}{d\Ptm},
  \label{eq:compliance}
\end{equation}
whose slope $dr_e/d\Ptm$ follows from the wall law (Eq.~\ref{eq:wall}). Each
edge splits its $C_e$ equally between its two endpoint nodes, so the nodal
compliance is the sum over the incident edges, $C_v=\sum_{e\ni v}\tfrac12\,C_e$.
This wall storage enters the pressure solve implicitly through the nodal system
(Eq.~\ref{eq:instep}, Sec.~\ref{sec:pipeline}).

\paragraph{Dynamic valve gating.}
The secondary bicuspid valves are modelled, as in the static study \cite{Chang2026Static}, by
distension-dependent thresholds on the signed transvalvular pressure
$\Delta p = p_i-p_j$ \cite{Davis2011,Wilson2015,Bertram2014},
\begin{align}
  \Delta p_{\mathrm{close}}^{(e)} &= a_p^{(e)}\exp(b_p^{(e)}\,\mathtt{x}),
    \label{eq:vclose}\\
  \Delta p_{\mathrm{open}}^{(e)}  &= c_p^{(e)}\!\left[1-\exp\!\big(d_p^{(e)}(\mathtt{x}-\mathtt{x}_0)\big)\right],
    \label{eq:vopen}
\end{align}
where $\mathtt{x}=r/r_{\max}$ is the fractional distension, with $r_{\max}$ the maximum distended radius of that sub-edge. The valve resistance
factor follows a state-latched sigmoid between the open and closed extremes
$\phi\in[0,370]$ \cite{Bertram2014,Bertram2020},
\begin{equation}
  \phi(\Delta p,s) = \phi_{\mathrm{open}} +
    \frac{\phi_{\mathrm{closed}}-\phi_{\mathrm{open}}}
    {1+\exp\!\big[\sigma_{\mathrm{inv}}^{(e)}(\Delta p-\Delta p_s^{(e)})\big]},
  \label{eq:sigmoid}
\end{equation}
with $s\in\{\mathrm{open},\mathrm{close}\}$ latched by a dead-band.
Unlike the static solve, the valve state is now advanced \emph{dynamically}: at
each step the per-valve $\Delta p$ is evaluated, the latch flips when the
threshold band is crossed, and $\phi$, hence the edge conductance in
Eq.~(\ref{eq:poiseuille}), is updated before the pressure solve. The sigmoid
steepness $\sigma_{\mathrm{inv}}^{(e)}$ is rescaled to human pressure units
(Sec.~\ref{sec:species}), and the rat literature value
($\sigma_{\mathrm{inv}}^{(e),\mathrm{rat}}=0.1$\,Pa$^{-1}$) seeds it before it is set
to an operating point as follows.

We keep Eqs.~(\ref{eq:vclose})--(\ref{eq:vopen}) in their general form but
set the edge-wise parameters $a_p^{(e)}=c_p^{(e)}=0$ throughout this study, for simplicity. Both thresholds then
vanish, so the dead-band closes and the sigmoid of Eq.~(\ref{eq:sigmoid}) centres
on $\Delta p=0$. Each gate then switches on the sign of $\Delta p$ with no
memory, and the only parameter left is the transition width, which we calibrate
per gate valve in Eq.~(\ref{eq:sigmainv_beta}) below. Restoring the full thresholds and
calibrating $\phi$ edge by edge is the subject of a companion study.

\paragraph{Operating-point calibration of the valve steepness.}
The sigmoid steepness is set per gate valve, not as one global value, so
each valve keeps the same relative transition width in $\Delta p$,
\begin{equation}
  \sigma_{\mathrm{inv}}^{(e)} \;=\; \frac{\beta}{\max\!\big(|\Delta p_e^{0}|,\,\Delta p_{\min}\big)},
  \label{eq:sigmainv_beta}
\end{equation}
where $\Delta p_e^{0}$ is the quasi-static (passive, $\Tact=0$) baseline
transvalvular pressure at gate edge $e$, we assume $\beta$ as a single dimensionless
sharpness shared by all valves (its value is given in Sec.~\ref{sec:results}), and $\Delta p_{\min}$ is a denominator floor for numerical stability. Away from that floor each valve has the same relative transition width
$1/\beta$, independent of its local pressure scale.

\paragraph{Valve placement and gating.}
Each valve sits at an interior lymphangion boundary and gates on a local pressure
difference. There is one valve per interior vertex
created by the lymphangion subdivision, while the original graph vertices
(branch points and terminal leaves) carry no valve. Each lymphangion is further resolved into
$N_g$ wall sub-segments (see figure~1 in \cite{Chang2026Static}), and the transvalvular pressure gating a valve is read
from the sub-segment pressures straddling it (the subscript $v$ here
labels the valve, not the lymphovenous-junction node of Sec.~\ref{sec:outlet}),
\begin{equation}
  \Delta p_v = \sum_{k} w_k\,\big(p_{\mathrm{up},k}-p_{\mathrm{dn},k}\big),
  \qquad w=(0.6,0.3,0.1),
  \label{eq:dpvalve}
\end{equation}
a weighted difference over the three nearest sub-vertices on each side
($k=1,2,3$ counted outward from the valve, pairing the upstream and downstream
sub-vertex at each offset). The per-valve state $s_v$ is advanced each step from
this $\Delta p_v$ and applied to both gate sub-segments.

Because the valve law $\phi$ describes a whole-lymphangion resistance while
the mesh subdivides each lymphangion into $N_g$ sub-segments, the per-valve
increment is scaled as $\phi\to\phi\,N_g/2$ across the two gate sub-segments, so
their series sum reproduces the intended lymphangion-level resistance.

\paragraph{Topology-ambiguous valves:}
A few valves ($\ll 1\%$) lie on edges whose steady flow direction is not fixed by the network
topology, and there the open or closed choice is immaterial. Such an edge carries
a transvalvular drop that hovers at noise level ($|\Delta p_v|\!\to\!0$), so its
flow direction is not resolved by the network geometry and the valve's
rectification orientation is itself ill-defined. Because the edge carries
negligible flow in either state, pinning it open or closed does not change the global solution. We thus exclude them from our statistical analysis. 

\subsection{Ostial valve outlet}
\label{sec:outlet}

The thoracic duct terminates at the lymphovenous junction (LVJ), where it drains
into the left subclavian--internal-jugular venous angle through an \emph{ostial
valve}, structurally and functionally distinct from the intraluminal
secondary valves of Sec.~\ref{sec:valve_dyn} \cite{Ratnayake2018}. Its two large,
obliquely set flap cusps are held open or closed primarily by the venous wall
tension, and hence by respiration (open on inspiration as the central-venous
pressure falls, closed on expiration), not by the transvalvular lymph pressure.
We therefore model the outlet as a respiration-gated two-state switch of the
ostial valve Eq.(\ref{eq:qout_gated}), not as the other lymph valves as in Eq.(\ref{eq:sigmoid}).

For the static outlet we reuse the ostial-valve model of the companion
paper \cite{Chang2026Static}. That model is a three-element (RCR) Windkessel in
its direct-current limit, which time-averages the respiration cycle into one
series resistance $R_{\mathrm{LVJ}}=1.96\times10^{10}$\,Pa\,s\,m$^{-3}$ to a
venous far field $P_{\mathrm{subclav}}=400$\,Pa ($3$\,mmHg). Its
construction and calibration are given there, and we retain it here as the static
reference and the cold-start initial condition for the dynamic runs.

Its calibration target, $\bar p_v\approx 8-10$\,mmHg at $2-4$\,L/day, is
porcine. It is the $8.2\pm3.2$\,mmHg control group of Lu et al.~\cite{Lu2022},
measured at the lympho-venous junction of domestic pigs weighing $55\pm7$\,kg. A
resting human junction pressure is not firmly established, and the direct human
recording of $16\pm5$\,mmHg \cite{Kelly2022} was made in patients under
positive-pressure mechanical ventilation.

The dynamic model replaces the ostial valve sigmoid with a two-state respiration switch
of the ostial valve. The open phase drains through an open-state resistance
$R_{\mathrm{open}}$, a free parameter in the low-resistance range of the sigmoid
(resolved below), and the closed phase is the sealed limit
($Q_{\mathrm{out}}=0$). The sigmoid therefore enters not as a continuous
instantaneous resistance but only through these two respiration-selected end
states. We reduce the full sigmoid to these two end states and do not fit
its continuous transition, because the ostial valve lacks the pressure-flow
data needed to constrain the sigmoid's parameters. This leaves a single free
parameter, the open-state resistance $R_{\mathrm{open}}$, which we fix as described
below. Respiration modulates thoracic-duct pressure and flow
\cite{Moazzam2026}, and the gate is what carries that modulation here. The
breathing cycle has
period $T_{\mathrm{resp}}=60/f_{\mathrm{resp}}$, set by the normal adult resting
rate $f_{\mathrm{resp}}\approx12$--$20$ breaths per minute
($T_{\mathrm{resp}}\approx3$--$5$\,s) \cite{Nicolo2020}. It alternates the
outlet between two boundary conditions,
\begin{equation}
  Q_{\mathrm{out}}(t)=
  \begin{cases}
    \dfrac{p_v(t)-P_{\mathrm{subclav}}}{R_{\mathrm{open}}},
      & I_{\mathrm{insp}}(t)=1 \ \text{and}\ p_v\ge P_{\mathrm{subclav}},\\[4pt]
    0, & \text{otherwise},
  \end{cases}
  \label{eq:qout_gated}
\end{equation}
where $p_v(t)$ is the LVJ nodal pressure. The inspiration indicator
$I_{\mathrm{insp}}(t)=1$ when
$\mathrm{mod}(t,T_{\mathrm{resp}})/T_{\mathrm{resp}}<f_{\mathrm{open}}\approx1/3$.

Here $f_{\mathrm{open}}$ is the inspiratory duty cycle
$T_{\mathrm{I}}/T_{\mathrm{tot}}$, which is measured directly in humans. Awake
resting recordings in $40$ healthy adults give
$T_{\mathrm{I}}/T_{\mathrm{tot}}=0.29\pm0.09$ \cite{Frizzelli2022}, so the value
$1/3$ used here sits inside one standard deviation of the measured duty cycle.

No direct measurement of the ostial-valve open resistance exists, in
human or in animal. The nearest pressure--flow data \cite{PatelLuKassab2025} come
from cannulated single-valve swine thoracic-duct segments perfused \emph{ex vivo}.
That valve is an intraluminal \emph{secondary} valve, structurally and
functionally distinct from the terminal ostial valve \cite{Ratnayake2018}, so it
fixes only an order of magnitude.

We therefore treat $R_{\mathrm{open}}$ as a free parameter, bounded below
by that secondary-valve open range ($\sim10^{8}$\,Pa\,s\,m$^{-3}$) and resolved by
a dynamic sweep. The sweep requires the breath-averaged outlet flow to reproduce
the physiological delivery and mean junction pressure, while the gating produces a
respiratory junction-pressure swing of order $1$\,mmHg \cite{Ratnayake2018}. It is
the instantaneous open-valve resistance, not the breath-averaged
$R_{\mathrm{LVJ}}$ of the static outlet \cite{Chang2026Static}, and is necessarily
smaller, since reusing $R_{\mathrm{LVJ}}$ would suppress that swing.

Because the ostial valve is anatomically one-way \cite{Ratnayake2018},
the closed branch is imposed whenever the junction-to-vein gradient reverses on
inspiration, so $Q_{\mathrm{out}}\ge0$ holds throughout.

On expiration the outlet carries no flux:
$Q_{\mathrm{out}}=0$, a homogeneous Neumann condition at the LVJ node. The rising
intrathoracic and central-venous pressure closes the valve and seals the duct
against venous reflux. Lymph still arrives from below but cannot leave. The
terminal-duct pressure is not pinned and rises freely over the expiratory
phase. The junction pressure swings over the breath cycle about its
time-averaged value. This respiration-gated outlet is the boundary condition used
in the dynamic runs reported below.

Because of breathing we also apply a pressure on the cisterna chyli, as in
the companion static model \cite{Chang2026Static}. This is a steady directional
respiratory drive of $\sim15$\,cmH$_2$O along the cisterna-chyli
edges, oriented toward the duct. It enters the dynamic runs as a fixed baseline and
is distinct from the live respiration-gated ostial outlet, so the two are not
double-counted.

The closed phase is a no-flux elastic reservoir (Neumann), not a
fixed-pressure sink, and it is well posed only because the terminal duct is
elastic. With $Q_{\mathrm{out}}=0$ the LVJ row of the in-step system
(Eq.~\ref{eq:instep}) carries no outlet drain and reduces to the local mass
balance
\begin{equation}
  C_v\,\frac{dp_v}{dt} \;+\; \sum_{j} g_{vj}\,(p_v-p_j)
    \;=\; Q_{\mathrm{src},v},
  \label{eq:closed_hold}
\end{equation}
where $C_v$ is the nodal wall compliance (the node-wise assembly of the
per-edge compliance $C_e$ of Eq.~\ref{eq:compliance}), $g_{vj}$ the Hagen--Poiseuille conductance to neighbour
$j$ at pressure $p_j$, and $Q_{\mathrm{src},v}$ the swept-volume source at the
junction. Lymph delivered by
$Q_{\mathrm{src},v}>0$ cannot leave and, since the compliant neighbours charge
together, the conduction term stays small ($p_v\sim p_j$), so the junction integrates its source,
$dp_v/dt\approx Q_{\mathrm{src},v}/C_v>0$: the sealed segment distends its own
wall and the pressure rises. The outlet node therefore \emph{retains} its wall
compliance $C_v$ and swept-volume source $Q_{\mathrm{src},v}$ in the closed
phase. %

\subsection{Time-stepping and solution procedure}
\label{sec:pipeline}

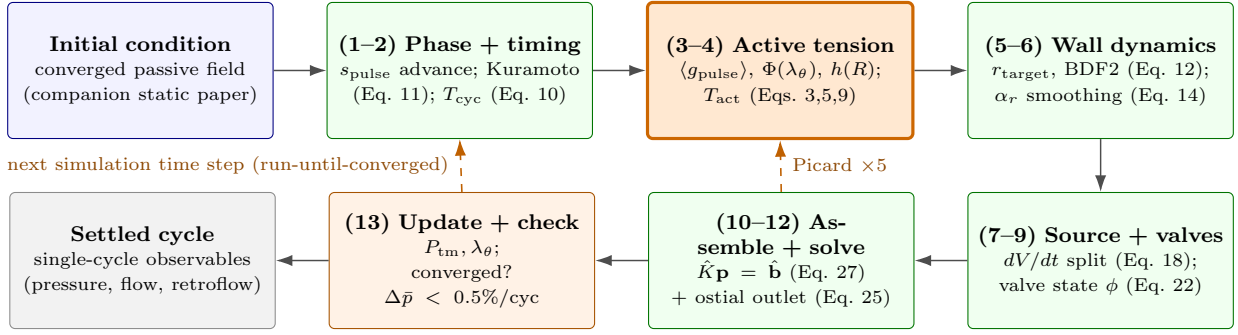
\begin{figure*}[t]
  \centering
  \begin{tikzpicture}[
    >={Latex[length=2.2mm,width=1.5mm]},
    node distance=7mm and 7mm,
    stage/.style={draw, rounded corners=2pt, align=center,
                  text width=33mm, minimum height=18mm, inner sep=3pt,
                  font=\footnotesize},
    input/.style={stage, fill=blue!5, draw=blue!50!black},
    solve/.style={stage, fill=green!6, draw=green!45!black},
    result/.style={stage, fill=orange!10, draw=orange!65!black},
    focus/.style={stage, fill=orange!18, draw=orange!80!black, very thick},
    valid/.style={stage, fill=gray!10, draw=black!45},
    flow/.style={->, line width=0.55pt, draw=black!70, rounded corners=4pt},
    loop/.style={->, line width=0.6pt, draw=orange!75!black, dashed, rounded corners=4pt}
  ]
    \node[input] (s0) {\textbf{Initial condition}\\[1pt]
      {\scriptsize converged passive field\\(companion static paper)}};
    \node[solve, right=of s0] (s1) {\textbf{(1--2) Phase + timing}\\[1pt]
      {\scriptsize $s_{\mathrm{pulse}}$ advance; Kuramoto\\(Eq.~\ref{eq:kuramoto});
       $\Tcyc$ (Eq.~\ref{eq:omega})}};
    \node[focus, right=of s1] (s2) {\textbf{(3--4) Active tension}\\[1pt]
      {\scriptsize $\langle\gpulse\rangle$, $\Phi(\lambda_\theta)$, $h(R)$;\\
       $\Tact$ (Eqs.~\ref{eq:Tact},\ref{eq:phi},\ref{eq:gpulse})}};
    \node[solve, right=of s2] (s3) {\textbf{(5--6) Wall dynamics}\\[1pt]
      {\scriptsize $r_{\mathrm{target}}$, BDF2 (Eq.~\ref{eq:wallode});\\
       $\alpha_r$ smoothing (Eq.~\ref{eq:alphar})}};
    \node[solve, below=of s3] (s4) {\textbf{(7--9) Source + valves}\\[1pt]
      {\scriptsize $dV/dt$ split (Eq.~\ref{eq:src});\\
       valve state $\phi$ (Eq.~\ref{eq:sigmoid})}};
    \node[solve, left=of s4] (s5) {\textbf{(10--12) Assemble + solve}\\[1pt]
      {\scriptsize $\hat K\mathbf{p}=\hat{\mathbf{b}}$ (Eq.~\ref{eq:instep})\\
       $+$ ostial outlet (Eq.~\ref{eq:qout_gated})}};
    \node[result, left=of s5] (s6) {\textbf{(13) Update + check}\\[1pt]
      {\scriptsize $\Ptm,\lambda_\theta$;\\
       converged? $\Delta\bar p<0.5\%$/cyc}};
    \node[valid, left=of s6] (s7) {\textbf{Settled cycle}\\[1pt]
      {\scriptsize single-cycle observables\\(pressure, flow, retroflow)}};

    \draw[flow] (s0) -- (s1);
    \draw[flow] (s1) -- (s2);
    \draw[flow] (s2) -- (s3);
    \draw[flow] (s3) -- (s4);
    \draw[flow] (s4) -- (s5);
    \draw[flow] (s5) -- (s6);
    \draw[flow] (s6) -- (s7);
    \draw[loop] (s5.north) -- node[right=1pt, font=\scriptsize, text=orange!55!black]
      {Picard $\times5$} (s2.south);
    \draw[loop] (s6.north) -- node[left=1pt, font=\scriptsize, text=orange!55!black]
      {next simulation time step (run-until-converged)} (s1.south);
  \end{tikzpicture}
  \caption{Dynamic solution procedure (Sec.~\ref{sec:pipeline}). The converged
  passive field of the companion static study is the initial condition. Each simulation
  step advances the contraction phase and cycle timing, assembles the active
  tension, advances the wall and sub-segment dynamics, computes the swept-volume
  source and valve state, and re-solves the nodal hydraulic system with the
  ostial-valve outlet. The intra-step active-tension/wall/pressure
  coupling is iterated to a fixed point by the Picard inner loop (short dashed
  return). The whole step repeats until the settled-cycle criterion of
  Sec.~\ref{sec:pipeline} is met (long dashed return). Box numbers
  match the step list in Sec.~\ref{sec:pipeline}.}
  \label{fig:pipeline}
\end{figure*}

Starting from the converged passive field as the initial condition, each simulation time
step $t^n\to t^{n+1}$ executes the following sequence (Fig.~\ref{fig:pipeline}):
(1) advance the per-lymphangion phase clock $s_{\mathrm{pulse}}$ (and, if
coupling is on, the Kuramoto phases), (2) update $\Tcyc$ via
Eq.~(\ref{eq:omega}) if pressure--frequency coupling is enabled, (3) evaluate the
chain-averaged $\langle\gpulse\rangle$ (Eq.~\ref{eq:gpulse}), (4) assemble the
active tension $\Tact$ (Eq.~\ref{eq:Tact}), (5) form the radius target from the
clamped Laplace balance, (6) advance the wall ODE (Eq.~\ref{eq:wallode}, BDF2) and smooth the
sub-segment radii (Eq.~\ref{eq:alphar}),
(7) compute the swept-volume source $dV/dt$ and split it (Eq.~\ref{eq:src}), (8)
recompute the lumped wall compliance, (9) update the valve state machine and
$\phi$ (Eqs.~\ref{eq:vclose}--\ref{eq:sigmoid}), (10) rebuild the edge
conductances, (11) assemble the nodal system with all sources, sinks, and the
ostial-valve outlet (Eq.~\ref{eq:qout_gated}),
(12) solve the sparse linear system
$\hat K\,\mathbf{p}^{n+1}=\hat{\mathbf{b}}$, and (13) update $\Ptm$,
$P_{\mathrm{passive}}$, and $\lambda_\theta$ for the next step.

The in-step nodal system is
\begin{equation}
  \left(K_{\mathrm{cond}} + \tfrac{C_v}{\Delta t}\,I\right)\mathbf{p}^{n+1}
    = \mathbf{Q}_{\mathrm{src}}^{n+1} + \tfrac{C_v}{\Delta t}\,\mathbf{p}^{n}
    + \mathbf{b}_{\mathrm{BC}},
  \label{eq:instep}
\end{equation}
solved by sparse LU factorization, where $K_{\mathrm{cond}}$ is the
Hagen--Poiseuille conductance matrix \cite{Savinkov2020}, $I$ the identity, $C_v/\Delta t$ the
implicit wall compliance on the diagonal, and $\mathbf{b}_{\mathrm{BC}}$ the
boundary vector holding the ostial-valve outlet and the imposed leaf inflows.
Step (12) above writes Eq.~(\ref{eq:instep}) compactly as
$\hat K\,\mathbf{p}^{n+1}=\hat{\mathbf{b}}$.
The wall-storage flux $C_e\,d\Ptm/dt$ (Eq.~\ref{eq:compliance}) is
discretized by backward Euler, which places $C_e/\Delta t$ on the diagonal and
$C_e\,p^{n}/\Delta t$ on the right-hand side at the edge endpoints, assembled
node-wise into the diagonal $C_v$. Steps (3)--(8) are wrapped in the per-step
Picard loop of Sec.~\ref{sec:wallode}, and all adopted parameter values are
collected in Table~\ref{tab:params}.

The settled state is a \emph{nonlinear limit cycle}, not a static fixed point.
At its operating point the system is linearly stable (the empirical Jacobian of the
step-to-step map of the cycle-averaged pressure has spectral radius $<1$ for
coherent, single-cell, and grid-frequency perturbations), so the sustained
oscillation is driven by the finite-amplitude phasic forcing rather than by a
linear instability. We therefore advance the integration until this state
is reached rather than for a fixed cycle count: the run stops once the
cycle-averaged nodal pressure changes by less than $0.5\%$ per cycle for three
consecutive cycles after a four-cycle warm-up, with a hard cap of $52$ cycles.
Cycle-resolved observables are then measured on one to three settled
cycles as defined in Sec.~\ref{sec:observables}.

\subsection{Species rescaling}
\label{sec:species}

The contraction parameters drawn from rodent measurements are mapped to human
scales by the same dimensional rescaling used in the static study \cite{Chang2026Static}. The phasic-tension and
pressure-sensitivity scales use the mouse$\to$human pressure ratio
$\alpha_p^{\mathrm{mouse}}$ ($\sphase^{\mathrm{human}}=\sphase^{\mathrm{mouse}}\,
\alpha_p^{\mathrm{mouse}}$, Eq.~\ref{eq:sigmaphase},
$\mu_p^{\mathrm{human}}=\mu_p^{\mathrm{mouse}}/\alpha_p^{\mathrm{mouse}}$). The $\mu_p$
cascade uses the production ratio $\alpha_p^{\mathrm{mouse}}=4.39$, which gives the
$4.74\times10^{-3}$\,Pa$^{-1}$ of Table~\ref{tab:params}. The $3.3$ listed in
Table~\ref{tab:params2} is the literature anchor of the $\sphase$ reference alone.
The valve threshold scales $a_p,c_p$ use the rat ratio $\alpha_p^{\mathrm{rat}}$
with the sigmoid steepness scaling inversely,
$\sigma_{\mathrm{inv}}^{\mathrm{human}}=\sigma_{\mathrm{inv}}^{\mathrm{rat}}/\alpha_p^{\mathrm{rat}}$,
so that the exponent $\sigma_{\mathrm{inv}}\Delta p$ in Eq.~(\ref{eq:sigmoid})
stays dimensionless under the pressure rescale.
The dimensionless shape parameters ($b_p,d_p,\mathtt{x}_0$,
$\Ton/\Tcyc$) are unchanged. %

\begin{table*}[t]
  \centering
  \caption{Dynamic-model parameter set (production values) needed to reproduce a
  run, part one, the pump itself. Table~\ref{tab:params2} carries the transport
  path from the valves to the venous outlet. Rows grouped by model component.
  Equation/table references point to the defining relation. Species ratios are
  literature anchors (Sec.~\ref{sec:species}). The globally constant $\sphase$ is
  swept (Eq.~\ref{eq:sigmaphase}), with literature anchor $5445$\,Pa
  ($\alpha_p^{\mathrm{mouse}}=3.3$) and a historical low anchor $4851$\,Pa
  ($=1650\times2.94$).}
  \label{tab:params}
  \small
  \begin{tabular}{llll}
    \toprule
    Symbol & Value & Units & Description / source \\
    \midrule
    \multicolumn{4}{l}{\textit{Discretization and numerics}}\\
    $N_g$ & 5 & dim &wall sub-segments per lymphangion \\
    $s_{\mathrm{valve}}$ & 3 & mm & inter-lymphangion (valve) spacing \cite{MargarisBlack2012} \\
    $\Delta t$ & 0.1 & s & pump time step \\
    cycles & $\le52$ (run-until-converged) & dim &stop at $<0.5\%$/cyc $\times3$ (Sec.~\ref{sec:pipeline}) \\
    $\eta$ & $1.8\times10^{-3}$ & Pa\,s & lymph dynamic viscosity \\
    \midrule
    \multicolumn{4}{l}{\textit{Active tension (Eq.~\ref{eq:Tact})}}\\
    $\sphase$ & 4851--100000 (sweep; lit 5445) & Pa & phasic-tension scale, globally constant per run \\
    $a_{\mathrm{tone}}$ & 0 & dim &tonic coefficient ($\stone$ disabled) \\
    $\lambda_0,\lambda_M$ & 0.25, 1.10 & dim &stretch window (Eq.~\ref{eq:phi}) (formula is from \cite{Razavi2020}) \\
    $\lambda_\theta$ divisor & 4 & dim &$\lambda_\theta=r/(4r_c)$ (Eq.~\ref{eq:lambdatheta}) \\
    $h$ intercept, slope & 37.0, 8.3 & $\mu$m, $\mu$m/mm & wall thickness (Table~\ref{tab:hwall}) \\
    $f_{\mathrm{SNP}}$ & 1 & dim &NO-mediated relaxation (disabled) \\
    $g_{\mathrm{spatial}}$ & $2\sin^2$ (or uniform) & dim &axial force profile (choice has minor effect) \\
    active lymphangions & interior only & dim &end lymphangions passive ($\Tact=0$) \\
    \midrule
    \multicolumn{4}{l}{\textit{Passive wall law (substrate, paper-form)}}\\
    $k_1$ & $1.569\times10^{-5}$ & Pa\,m & Athanasiou paper-form fit \cite{Athanasiou2017} \\
    $k_2$ & 4.31 & dim &Athanasiou paper-form fit \cite{Athanasiou2017} \\
    $P_{\mathrm{anchor}}^{\mathrm{paper}}$ & 490.33 & Pa & $5$\,cmH$_2$O reference \cite{Athanasiou2017} \\
    \midrule
    \multicolumn{4}{l}{\textit{Wall dynamics}}\\
    $\tau_{\mathrm{wall}}=b/k$ & 0.333 & s & over-damped visco-elastic relaxation (Eq.~\ref{eq:wallode}); value $\sim c_2^{-1}$ \cite{Contarino2017} \\
    $\tau_f$ & 0.333 & s & $r_{\mathrm{target}}$ filter, $\equiv\tau_{\mathrm{wall}}$ (band-pass) \\
    $\gamma$ & 0.3 & dim &$=\Delta t/\tau_{\mathrm{wall}}$ (BDF2) \\
    $\alpha_r$ & 0.2 & dim &sub-segment Laplacian smoothing (Eq.~\ref{eq:alphar}) \\
    $N_{\mathrm{Picard}}$ & 5 & dim &per-step Picard sweeps \\
    fluid inertance & off & dim &$L_h\,dQ/dt$, negligible in periphery, TD $W_o^{(e)}\!\approx\!1$ phase-lag limit \\
    \midrule
    \multicolumn{4}{l}{\textit{Contraction cycle / waveform}}\\
    $f_{\mathrm{terminal}}$ & per region & s$^{-1}$ &terminal frequency (Table~\ref{tab:tcyc}) \cite{Zawieja2018} \\
    $\Ton/\Tcyc$ & 0.293 & dim &duty ratio \cite{Davis2025} \\
    $\mu_p$ & $4.74\times10^{-3}$ & Pa$^{-1}$ & P-frequency coupling (Eq.~\ref{eq:omega}; off at baseline) \\
    \midrule
    \multicolumn{4}{l}{\textit{Contraction phase}}\\
    $\Delta\phi$ & 0 & deg & initial phase lag (in-phase) \\
    $K_{\mathrm{couple}}$ & 0 & rad/s & coupling off (Eq.~\ref{eq:kuramoto}); coupled study in companion paper \\
    \bottomrule
  \end{tabular}
\end{table*}

\begin{table*}[t]
  \centering
  \caption{Dynamic-model parameter set, part two, the transport path from the
  valves to the venous outlet. Conventions as in Table~\ref{tab:params}.}
  \label{tab:params2}
  \small
  \begin{tabular}{llll}
    \toprule
    Symbol & Value & Units & Description / source \\
    \midrule
    \multicolumn{4}{l}{\textit{Valve mechanics (Eqs.~\ref{eq:vclose}--\ref{eq:sigmoid})}}\\
    $a_p^{\mathrm{rat}}$ & $2.30\times10^{-4}$ & cmH$_2$O & close-threshold scale \cite{Davis2011} \\
    $b_p$ & 9.04 & dim &close-threshold exponent \cite{Davis2011} \\
    $c_p^{\mathrm{rat}}$ & $3.24\times10^{-2}$ & cmH$_2$O & open-threshold scale \cite{Davis2011} \\
    $d_p$ & 14.70 & dim &open-threshold exponent \cite{Davis2011} \\
    $\mathtt{x}_0$ & 0.78 & dim &half-opening distension \cite{Davis2011} \\
    $\sigma_{\mathrm{inv}}^{\mathrm{rat}}$ & 0.1 & Pa$^{-1}$ & sigmoid steepness \cite{Bertram2020} \\
    $\phi_{\mathrm{open}},\phi_{\mathrm{closed}}$ & 0, 370 & dim &valve resistance extremes \cite{Bertram2014} \\
    valve placement & interior vertices & dim &one per lymphangion boundary \\
    $\Delta p_v$ weights & 0.6 / 0.3 / 0.1 & dim &3-nearest sub-vertex gate (Eq.~\ref{eq:dpvalve}) \\
    \midrule
    \multicolumn{4}{l}{\textit{Contraction-volume source (Eq.~\ref{eq:src})}}\\
    source split & 0.5 / 0.5 & dim &equal to both endpoints, contract / relax \\
    strength factor & 1.0 & dim &$dV/dt$ source multiplier \\
    leaf clamp & off & dim &inflow leaves do receive the wall source \\
    \midrule
    \multicolumn{4}{l}{\textit{Ostial valve outlet (Eq.~\ref{eq:qout_gated})}}\\
    $R_p$ & $4.495\times10^{9}$ & Pa\,s/m$^3$ & ostial RCR proximal \cite{Lu2022} \\
    $R_d$ & $1.509\times10^{10}$ & Pa\,s/m$^3$ & ostial RCR distal \cite{Lu2022} \\
    $R_{\mathrm{LVJ}}$ & $1.96\times10^{10}$ & Pa\,s/m$^3$ & effective resistance $=R_p+R_d$ \\
    $C$ & $10^{-15}$ & m$^3$/Pa & DC limit (mandatory) \\
    $P_{\mathrm{subclav}}$ & 400 & Pa & far field, $3$\,mmHg \cite{lloyd2025measurement} \\
    $R_{vl},R_{vh}$ & $\sim$1e8, $\le$6e10 & Pa\,s\,m$^{-3}$ & secondary-valve order-of-magnitude ref. \cite{PatelLuKassab2025} \\
    $s$ & 15--86 & cmH$_2$O$^{-1}$ & secondary-valve sigmoid steepness ref. \cite{PatelLuKassab2025} \\
    CC respiration & 1471 & Pa & cisterna-chyli respiratory pump (total) \\
    \midrule
    \multicolumn{4}{l}{\textit{Species rescaling (Sec.~\ref{sec:species})}}\\
    $\alpha_p^{\mathrm{rat}}$ & 9.1 & dim &human/rat pressure \cite{Lu2022} \\
    $\alpha_p^{\mathrm{mouse}}$ & 3.3 & dim &human/mouse pressure (lit $\sphase$ anchor) \cite{Lu2022} \\
    $\lambda_r^{\mathrm{rat}},\lambda_L^{\mathrm{rat}}$ & 4.34, 7.97 & dim &diameter, length ratios \\
    $\lambda_r^{\mathrm{mouse}},\lambda_L^{\mathrm{mouse}}$ & 4.48, 11.9 & dim &diameter, length ratios \\
    \midrule
    \multicolumn{4}{l}{\textit{Transmural leak and lymph-node retention}}\\
    $L_{\mathrm{leak}}$ & $1.25\times10^{-7}$ & cm/s/cmH$_2$O & wall filtration \cite{jannaway2021ve} \\
    $f_n$ & 0.10 & dim &lymph-node retention fraction \\
    $R_n$ & $7.65\times10^{10}$ & Pa\,s/m$^3$ & lymph-node resistance \cite{Jafarnejad2015} \\
    \bottomrule
  \end{tabular}
\end{table*}

\subsection{Pumping observables}
\label{sec:observables}

Cycle-resolved quantities are measured over one to three settled cycles at
the end of the run. The contraction pressure amplitude
$\Delta P_{\mathrm{contraction}}$, written $\Delta P$ below, at a node is the peak-to-trough difference
across that window, with the cycle period \emph{detected from
the signal} rather than assumed. A max$-$min over the whole run is
deliberately avoided, as it folds in startup transient and envelope drift and
overstates the amplitude.

\section{Results}
\label{sec:results}

The dynamic runs test whether the lymphovenous junction reaches a
respiration-gated pumping cycle. We report a few benchmark runs at
representative values of the phasic-tension scale $\sphase$ and the valve-sigmoid
sharpness $\beta$, not a full grid. Here $\beta$ is the valve-sigmoid sharpness of
Eq.~(\ref{eq:sigmainv_beta}), a numerical-constitutive sensitivity parameter
rather than a biological axis, and $\beta=10$ is the validation reference.

We report only the thoracic-duct lymphovenous junction as output node (LVJ), the terminal-duct observable, which is the single clinically measured
site. Each run advances the LVJ to a quasi-steady respiration-gated limit cycle.
We make no claim of full-network equilibrium. The runs are fixed-duration
($t=200$\,s, $\approx20$ contraction cycles), sufficient for the LVJ operating
point but not a whole-network steady state.

\subsection{Emergence of a quasi-steady pumping state}
\label{sec:res-pump}

The activated network settles from its static initial condition into a
quasi-steady, respiration-gated pumping cycle at the lymphovenous junction, and the
operating pressure and drainage it reaches lie in the physiological range.

\begin{figure}[t]
  \centering
  \includegraphics[width=\columnwidth]{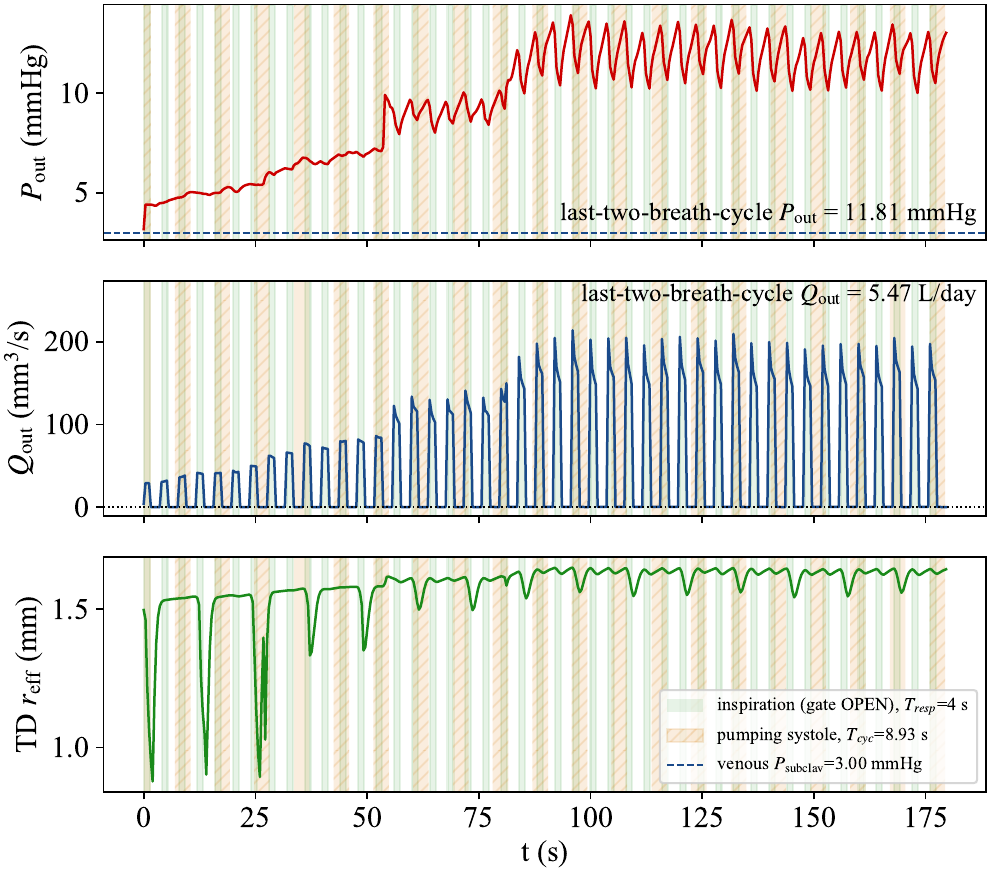}
  \caption{Emergence of a quasi-steady pumping state, benchmark run at
  $\sphase=30$\,kPa and $\beta=10$. From top to bottom: the LVJ outlet pressure
  $P_{\mathrm{out}}$, the gated outflow $Q_{\mathrm{out}}$, and the thoracic-duct
  radius $r_{\mathrm{eff}}$ against time. From rest the network charges over the
  first $50$ to $100$\,s and then holds a bounded limit cycle. The green band is
  the respiration inspiration window ($T_{\mathrm{resp}}=4$\,s, ostial gate open)
  and the orange hatching is the pumping systole ($T_{\mathrm{cyc}}\approx9$\,s).
  The dashed line is the venous reference.}
  \label{fig:pump}
\end{figure}

Figure~\ref{fig:pump} shows a benchmark run at $\sphase=30$\,kPa and
$\beta=10$, with the LVJ outlet pressure $P_{\mathrm{out}}$ ($\equiv p_v$ of
Eq.~(\ref{eq:qout_gated})), the gated outflow
$Q_{\mathrm{out}}$, and the thoracic-duct radius $r_{\mathrm{eff}}$ against time
from top to bottom. Turning on the contraction distends and
contracts the wall dynamically, so the lymph volume swings above and below its
static value. The outlet therefore first fills rather than drains, and
$P_{\mathrm{out}}$ charges over the first $50$ to $100$\,s toward a quasi-steady
operating pressure. Once the network reaches its dynamical equilibrium the
cycle-averaged pressure and outflow become nearly constant, and we run enough
breath cycles to confirm this convergence. The bottom trace is the predicted
physical thoracic-duct radius. At this $\sphase$ the contraction drives the
operating point up until the rising pressure balances the muscle, so
$r_{\mathrm{eff}}$ also reaches a dynamical equilibrium and stays nearly constant at LVJ, but changing of radius $dr$ is slightly larger at upstream of TD. This small $dr$ may imply that the $\sphase$ is too small, we thus plan a future work to investigate this.   

The charging transient follows from the network being an elastic
reservoir. Every wall in the graph is compliant, so the volume the network holds
is set by the transmural pressure through the wall law of
Sec.~\ref{sec:wallode}. At $t=0$ the walls sit at their static radii. Switching
on the contraction raises the mean transmural pressure, the wall radius distend, and
the network takes up extra fluid. Inflow exceeds outflow for as long as that
storage is filling, which is why $P_{\mathrm{out}}$ rises before the outlet
starts to drain.

The approach to that equilibrium is not complete at the end of a run. An
elastic reservoir relaxes on the timescale set by its own compliance, and that
timescale is long compared with one contraction cycle, so a small residual
remains after the outlet has reached its operating point. We measure the
residual over the last ten contraction cycles of every cell in the
$\sphase\times P_{\mathrm{subclav}}$ sweep. The network-mean wall distension
still moves by $0.04$ to $0.12$\% per cycle and the network-mean pressure by
$0.40$ to $0.51$\% per cycle. The reported $P_{\mathrm{out}}$ mean moves by $0.97$
to $2.30$\% between consecutive ten-breath windows.

Those residual is ignorable against the effects reported here. Across the
swept range $P_{\mathrm{out}}$ rises by a factor of $2.0$ to $2.4$ with
$P_{\mathrm{subclav}}$, the factor falling as $\sphase$ rises, and the spread
across $\sphase$ reaches $19$\% at the
lowest $P_{\mathrm{subclav}}$. Only the $\sphase$ spread at the two highest
$P_{\mathrm{subclav}}$ values falls to the size of the residual, and no
$\sphase$ claim is made there.

\begin{figure*}[t]
  \centering
  \includegraphics[width=\textwidth]{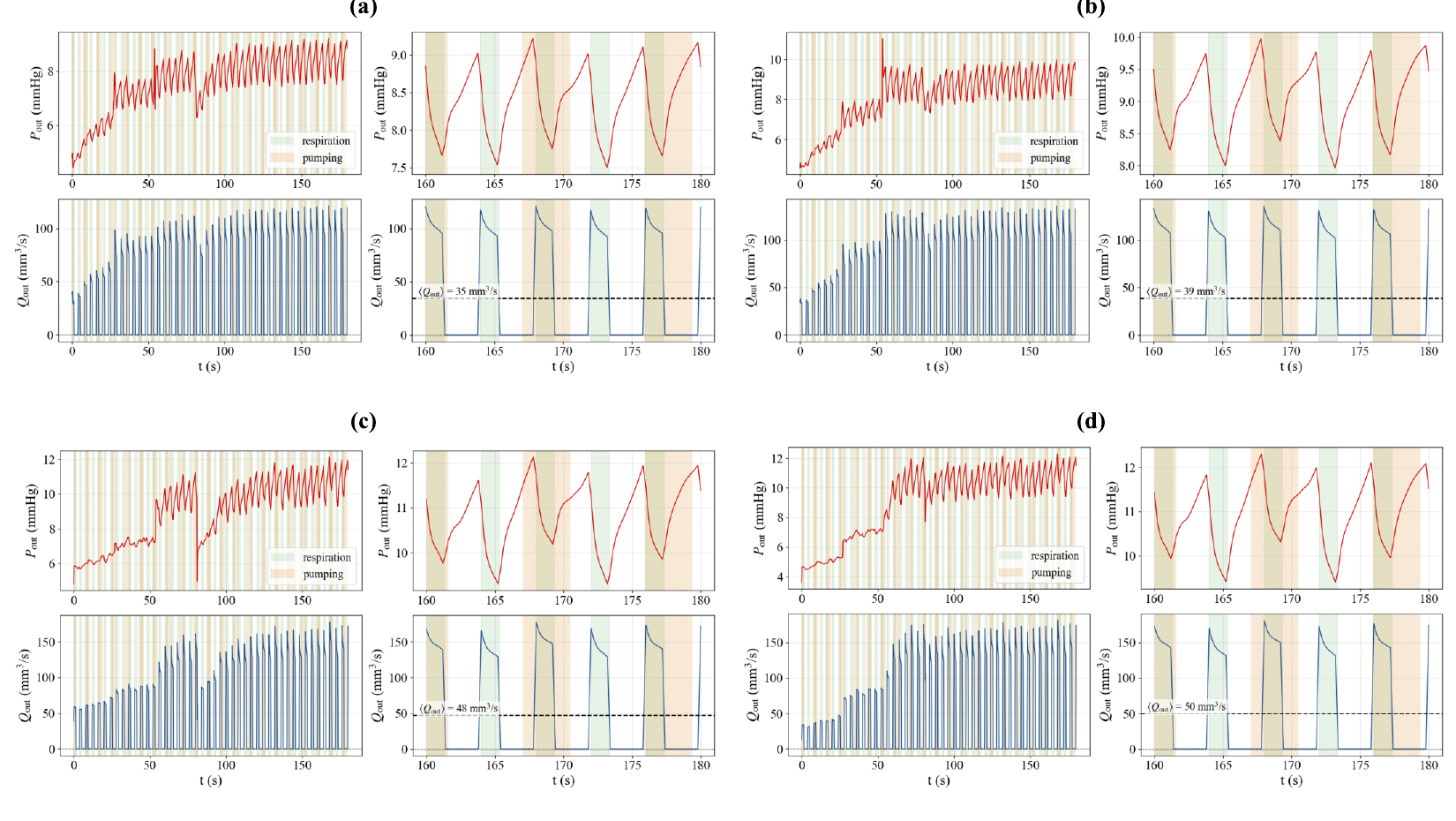}
  \caption{Representative pumping time series for four benchmark runs,
  spanning $\sphase=10$ to $30$\,kPa and $\beta=0.5$ to $10$. Each panel shows the
  LVJ outlet pressure $P_{\mathrm{out}}$ (top) and gated outflow $Q_{\mathrm{out}}$
  (bottom, the simulated ostial outlet flow of Eq.~\ref{eq:qout_gated}) over the
  full run (left) and the settled tail (right). Green marks the respiration
  inspiration window (ostial gate open) and orange hatching marks the pumping
  systole. (a)~$\sphase=10$\,kPa, $\beta=0.5$. (b)~$\sphase=15$\,kPa, $\beta=1$.
  (c)~$\sphase=30$\,kPa, $\beta=0.5$. (d)~$\sphase=30$\,kPa, $\beta=10$.}
  \label{fig:examples}
\end{figure*}

Figure~\ref{fig:examples} shows the full pumping dynamics behind the benchmark
summary. Each panel plots the LVJ outlet pressure $P_{\mathrm{out}}$ and the gated
outflow $Q_{\mathrm{out}}$ against time, over the whole run on the left and over the
settled tail on the right. The two shaded bands separate the two rhythms that drive
the outlet. The green band is the respiration inspiration window
($T_{\mathrm{resp}}=4$\,s), when the ostial gate is open. The orange-hatched band
is the pumping systole $T_{\mathrm{cyc}}=1/f_{\mathrm{cyc}}\approx9$\,s, when
the lymphangions contract ($T_{\mathrm{on}}\sim3$\,s).

The outflow is the rectified, respiration-gated drainage of the ostial outlet
(Eq.~\ref{eq:qout_gated}, Sec.~\ref{sec:outlet}): the gate opens only on
inspiration, a window of duty $f_{\mathrm{open}}\approx1/3$ within each breath, and
the one-way valve admits only forward flow. This piecewise law is why
$Q_{\mathrm{out}}$ in Fig.~\ref{fig:examples} drains in inspiration pulses and sits
at zero through expiration, while $P_{\mathrm{out}}$ keeps both rhythms because the
closed gate traps the contraction swing inside the duct.

The settled tail carries a finer structure. During the pumping systole
(orange) the pressure runs higher, so the offset between the pressure peaks and
dips with and without an active contraction is a candidate marker of lymphangion
contractile health. When the end of a breath meets a contraction, the pressure
rises relatively fast until the contraction ends.

The four benchmark runs of Fig.~\ref{fig:examples} span the two knobs and
follow a single monotonic trend. Their breath-averaged outflow is
$Q\approx35,\ 39,\ 48,\ 50$\,mm$^3$/s and their LVJ pressure range is
$7.5$--$9.4$, $8.0$--$10.0$, $9.4$--$12.1$, and $9.5$--$12.3$\,mmHg for panels
(a) $\sphase=10$\,kPa, $\beta=0.5$, (b) $\sphase=15$\,kPa, $\beta=1$, (c) $\sphase=30$\,kPa,
$\beta=0.5$, and (d) $\sphase=30$\,kPa, $\beta=10$, respectively. Both the pressure and the outflow rise
with $\sphase$, the signature of stronger active contraction driving a higher
operating point and greater net drainage. These benchmark pressures lie within the
measured terminal-duct range \cite{Lu2022}, a boundary-condition consistency check.

\subsection{Contraction pressure and flow amplitude}
\label{sec:res-amp}

\begin{figure}[t]
  \centering
  \includegraphics[width=\columnwidth]{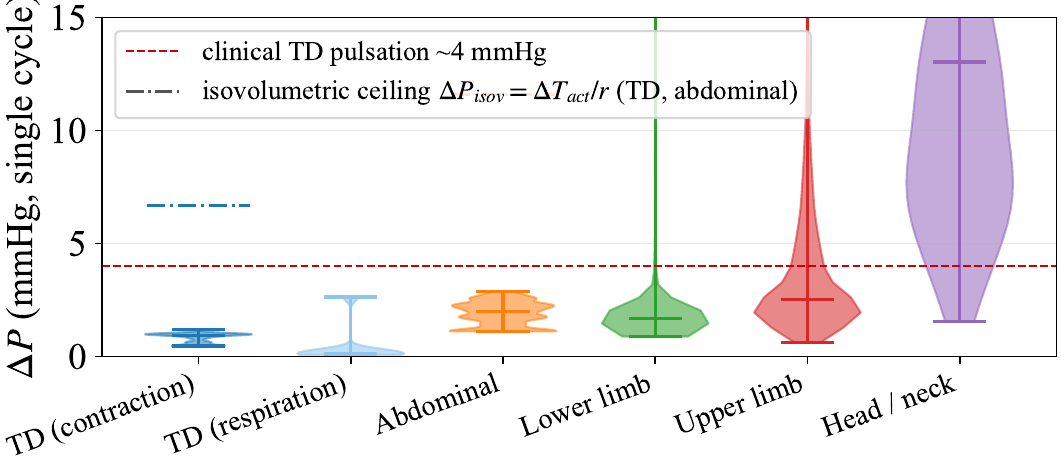}
  \caption{Per-region distribution of the single-cycle pressure amplitude
  $\Delta P$ across lymphangion subsegments. The thoracic-duct column is split
  into a contraction part (pump cycle, phase-folded at $T_{\mathrm{cyc}}$) and a
  respiration part (breathing gate, phase-folded at $T_{\mathrm{resp}}$). The four
  peripheral organs show the total swing, which is contraction-led there. Each
  violin shows the spread across subsegments with the median marked. The red
  dashed line is the clinical thoracic-duct pulsation ($\lesssim \,$4\,mmHg) and the
  dash-dot lines mark the per-region isovolumetric ceiling
  $\Delta P_{\mathrm{isov}}=\Delta\Tact/r$.}
  \label{fig:amp}
\end{figure}

Figure~\ref{fig:amp} reports the per-region single-cycle pressure amplitude
$\Delta P$ across all lymphangion subsegments, against the clinical thoracic-duct
$\Delta P_{\mathrm{TD}}\lesssim4$\,mmHg \cite{Kelly2022} and the isovolumetric ceiling.
The amplitude is set by the vessel's resistance to compression. If a lymphangion
were fully sealed the pressure would rise to
$\Delta P_{\mathrm{isov}}=\Delta\Tact/r\approx11$\,mmHg, about $2.8\times$ the
clinical value and the largest the muscle alone can raise. But as the muscle
shortens the transmural pressure and the passive wall push back, so the radius
moves by only a few percent per cycle and the model realizes only $\sim$14\% of
this ceiling. This compression back-reaction is
the dominant limit on the amplitude, and it leaves many organ vessels hard to
squeeze small.

The comparison with the clinical value is an open question. Kelly
\emph{et~al.}\ \cite{Kelly2022} advance a catheter into the thoracic part of the
duct and read a pressure increase of $\lesssim4$\,mmHg (a wide range,
$1$--$12$\,mmHg) as the intrinsic contraction, but the reading depends on where the
pressure is taken. Close to the ostial valve the swing is unavoidably set by the
respiration-driven opening and closing of the valve, and our model produces a
comparable $2$--$4$\,mmHg there (see Fig.~\ref{fig:examples}). In the thoracic duct proper, where their catheter
sits, the ostial influence is weaker, so a swing of this size would have to come
from contraction, which our model predicts about $\sim 1$--$2$\,mmHg. If the clinical
increase is contraction-driven, our phasic-tension scale $\sphase$ is then too
small to raise that much pressure. The reported contraction period of about
$12$\,s (frequency $5$\,min$^{-1}$) \cite{Kelly2022} matches our pumping cycle in either reading.
These clinical values also come from ventilated patients with congenital heart
defects whose thoracic duct may be functionally altered, so they serve only as an
approximate reference.

The long tails of the lower-limb and upper-limb distributions in
Fig.~\ref{fig:amp} come from very small vessels whose flow direction the coarse
graph does not resolve, a geometry limitation that a finer resolution would reduce.

The thoracic-duct pulsation is contraction-led inside the lumen but
respiration-led at the junction. We split it into these two sources by phase-folding
the lumen pressure at the pumping period $T_{\mathrm{cyc}}$ and at the respiration
period $T_{\mathrm{resp}}$. Inside the lumen the contraction swing (median
$\sim1$\,mmHg) dominates the respiration swing (median $\sim0.15$\,mmHg), because
respiration enters only through the outlet valve and its ripple decays as it travels
upstream. At the junction itself the respiration swing is far larger
($1$--$2.5$\,mmHg), of the order of the clinical $\lesssim4$\,mmHg increase and
matching the near-ostial reading of the open question above.

\subsection{Length--tension window}
\label{sec:res-window}
\begin{figure}[t]
  \centering
  \includegraphics[width=1.02\columnwidth]{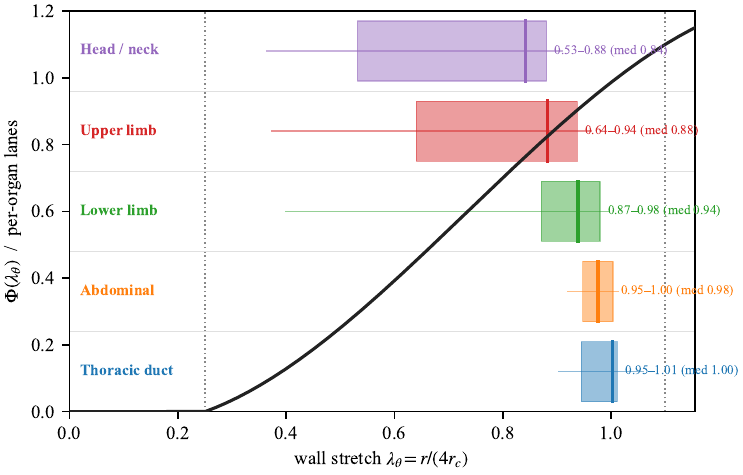}
  \caption{Active length-tension window $\Phi(\lambda_\theta)$
  Eq.~(\ref{eq:phi}) with stretch limits
  $\lambda_0=0.25$, $\lambda_M=1.10$ and the realized operating stretch drawn as
  one band per organ. The vertical axis $\Phi$ reads as the strength of the active tension and the
  horizontal axis $\lambda_\theta=r/(4r_c)$ tracks the radius, so a point further
  left is more compressed. Each coloured band is the operating-stretch distribution
  of one organ. The thoracic duct and the abdominal vessels sit at the highest
  stretch ($\lambda_\theta\approx0.98$), because their higher transmural pressure
  makes them hard to contract and keeps their radius large. All organs lie on the
  falling side of the window, so $\Phi$ acts as a compression brake.}
  \label{fig:phiwindow}
\end{figure}

All organs operate on the braking side of the active length-tension window.
Figure~\ref{fig:phiwindow} shows the window $\Phi(\lambda_\theta)$ of
Eq.~(\ref{eq:phi}) together with the operating stretch $\lambda_\theta=r/(4r_c)$
(the form of Eq.~(\ref{eq:lambdatheta}) with $\lambda_0=1/4$) realized across the
network, drawn as one band per organ. The thoracic duct operates near
$\lambda_\theta\approx0.98$, and across all organs the realized stretch spans
$\lambda_\theta\approx0.53$ to $1.01$. Every organ therefore sits below the window
peak at $\lambda_\theta=\lambda_M=1.10$, which the wall law places at
$r=4.4\,r_c$, outside the physical range. Across the physical range $\Phi$ falls
monotonically as the vessel compresses toward $r_c$, so the window acts as a
natural brake instead of a tension peak \cite{Zhang2007,Telinius2010}, and no separate maximum-tension cap is
required.

\subsection{Wall-relaxation timescale}
\label{sec:res-tau}

The local wall filter is mild at the production $\tau_{\mathrm{wall}}$, but
the gated outlet opens a failure window at slower wall relaxation. The wall
relaxation time $\tau_{\mathrm{wall}}=b/k$ (Eq.~\ref{eq:wallode}) sets how fast
the radius tracks its Laplace target and is distinct from the contraction period
$\Tcyc$.

Linearizing Eq.~(\ref{eq:wallode}) about the operating radius and driving the target
sinusoidally, $r_{\mathrm{target}}(t)=\bar r+\delta\hat r_{\mathrm{target}}\,e^{i\omega t}$
(with $i=\sqrt{-1}$), gives the wall radius response
$r(t)=\bar r+\delta\hat r\,e^{i\omega t}$ with
complex transfer function
\begin{equation}
  H(i\omega) \;\equiv\; \frac{\delta\hat r}{\delta\hat r_{\mathrm{target}}}
    \;=\; \frac{1}{1+i\omega\tau_{\mathrm{wall}}},
  \qquad \omega=\frac{2\pi}{\Tcyc},
  \label{eq:walltf_def}
\end{equation}
i.e.\ the ratio of the wall-radius oscillation to the radius-target drive at the
contraction frequency.

\begin{figure}[t]
  \centering
  \includegraphics[width=0.86\columnwidth]{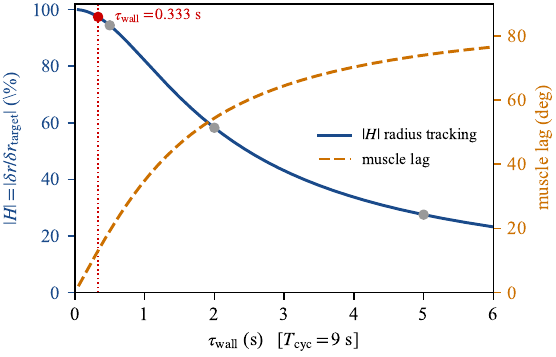}
  \caption{Wall-relaxation transfer function (Eq.~\ref{eq:walltf}): wall
  amplitude $|H|$ and muscle lag versus $\tau_{\mathrm{wall}}$ at $\Tcyc=9$\,s.
  Analysis in Sec.~\ref{sec:res-tau}.}
  \label{fig:tau_a}
\end{figure}

The magnitude and phase of $H$ are
\begin{equation}
  |H(\omega)| = \frac{1}{\sqrt{1+(\omega\tau_{\mathrm{wall}})^2}},
  \qquad
  \varphi(\omega) = -\arctan(\omega\tau_{\mathrm{wall}}),
  \label{eq:walltf}
\end{equation}
where $\varphi (\omega)$ is the wall-response phase lag, not the valve resistance factor
$\phi$. So at the production $\tau_{\mathrm{wall}}=0.333$\,s and $\Tcyc\approx9$\,s the
wall follows the target radius with $\approx97\%$ amplitude and a $13^\circ$ lag,
degrading to $58\%/54^\circ$ at $\tau_{\mathrm{wall}}=2$\,s and $28\%/74^\circ$ at
$5$\,s. Figure~\ref{fig:tau_a} plots this amplitude $|H|$ and the muscle
lag against $\tau_{\mathrm{wall}}$. At the production value
$\omega\tau_{\mathrm{wall}}\ll1$, so the wall tracks the contraction, while both the
amplitude falls and the lag grows as $\tau_{\mathrm{wall}}$ increases.

\begin{figure}[t]
  \centering
  \includegraphics[width=\columnwidth]{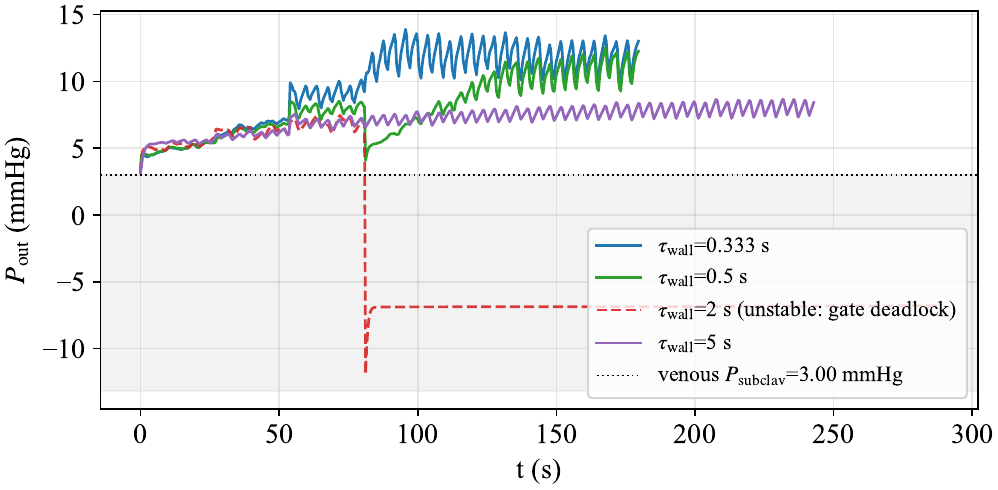}
  \caption{Simulated lymphovenous-junction pressure $P_{\mathrm{out}}(t)$ over
  a $\tau_{\mathrm{wall}}$ sweep ($\sphase=30$\,kPa, gated outlet). The
  $\tau_{\mathrm{wall}}=2$\,s case (dashed) collapses sub-venous. Analysis in
  Sec.~\ref{sec:res-tau}.}
  \label{fig:tau_b}
\end{figure}

The linear transfer function predicts a simple amplitude roll-off, but the
full nonlinear gated-outlet system does not degrade linearly at every
$\tau_{\mathrm{wall}}$. Figure~\ref{fig:tau_b} traces the lymphovenous-junction
pressure $P_{\mathrm{out}}(t)$ over a $\tau_{\mathrm{wall}}$ sweep at the headline $\sphase$.
The production $\tau_{\mathrm{wall}}=0.333$\,s and the $0.5$\,s case keep $P_{\mathrm{out}}$ above the
venous pressure throughout the breath and synchronized with the contraction.
The $\tau_{\mathrm{wall}}=5$\,s case remains stable at a lower operating point. At
$\tau_{\mathrm{wall}}=2$\,s, however, close to half the respiratory period
($T_{\mathrm{resp}}/2=2$\,s), a resonant interaction between the slow wall and
the respiration gate drives $P_{\mathrm{out}}$ transiently below $P_{\mathrm{subclav}}$, and the
one-way ostial valve then latches shut (closed in ${\sim}88\%$ of steps versus the
nominal $2/3$).

Once the valve is shut, the sealed terminal segment can regain pressure only
through its own wall contraction. With the outlet drain removed, the junction
obeys the closed-phase balance $C_v\,dp_v/dt=Q_{\mathrm{src},v}$
(Eq.~\ref{eq:closed_hold}), so its pressure rises only as fast as the swept-volume
source $Q_{\mathrm{src}}\propto dV/dt=2\pi r L\,(dr/dt)$ feeds it. A slow wall
relaxes slowly. The wall ODE $\tau_{\mathrm{wall}}\,dr/dt=r_{\mathrm{target}}-r$
gives a small $dr/dt$, so this source is weak. The wall then cannot lift the sealed
segment back above $P_{\mathrm{subclav}}$ before the next breath, the valve never
re-opens, and $P_{\mathrm{out}}$ settles at a sub-venous fixed point
(${\approx}-7$\,mmHg) where the outlet deadlocks (see the red-dashed line in Figure~\ref{fig:tau_b}). This failure is a
numerical-parameter instability window, not a physiological regime. It shows that
the small production $\tau_{\mathrm{wall}}$ keeps the gated outlet on its stable,
synchronized branch, and it bounds $\tau_{\mathrm{wall}}$ from above.

The wall timescale also invites an aging reading. In vivo, cervical
lymphatic vessels lose contractile function with age \cite{du2024restoration},
which in our framework would appear as a larger $\tau_{\mathrm{wall}}$ alongside a
smaller $\sphase$. A slower wall and a weaker drive lower the contraction and,
through the spread of per-lymphangion frequencies, make the pumping less ordered.

\subsection{Comparison with the static companion study}
\label{sec:res-companion}

Two quantities compare node for node with the static companion study,
which runs on the same graph at the same venous reference. Its lymphovenous
junction pressure at $v_{996}$ is $7.9$\,mmHg \cite{Chang2026Static}, against
$7.33$, $7.51$, $7.99$ and $8.82$\,mmHg here at $P_{\mathrm{subclav}}=400$\,Pa
for $\sphase=5445$, $10000$, $15000$ and $30000$\,Pa. Its delivered outflow is
$50.7$\% of a $65.750$\,mm$^3$/s whole-network gross inflow, or $33.3$\,mm$^3$/s, against
$33.65$\,mm$^3$/s here. Both agree within the spread that $\sphase$ alone
produces.

The four head and neck collectors of that study have dynamic
counterparts at the same vertices. It reports $10.2$, $13.6$, $15.5$ and
$17.6$\,mmHg at $v_{906}$, $v_{910}$, $v_{477}$ and $v_{474}$
\cite{Chang2026Static}. At the same venous reference and the weakest
contraction, $\sphase=5445$\,Pa, this model gives $11.8$, $13.2$, $18.8$ and
$26.2$\,mmHg. Raising $\sphase$ to $30000$\,Pa lifts them to $27.6$, $22.6$,
$41.3$ and $64.7$\,mmHg. The two collectors carrying the smallest static flow,
$v_{477}$ and $v_{474}$, gain the most, so contraction widens the pressure
spread across the territory by about a factor of three.

\section{Discussion}
\label{sec:discussion}

\subsection{Whole-body pumping and its clinical reading}
\label{sec:disc-pumping}

From first principles the network reaches a physiological pumping state.
With no fit to the dynamic data, the lymphovenous junction settles from a static
start into a bounded, respiration-gated cycle whose operating pressure and
drainage fall in the clinical range, and both rise with the phasic-tension scale
$\sphase$. A body-wide anatomical graph carrying contractile physics therefore
produces lymphatic pumping without hand-tuning, which no earlier model has shown
at this scale.

The model forces one conclusion that holds regardless of the unknown
parameters. A single globally constant contraction amplitude cannot reproduce the
whole body at once. The active tension has to be set vessel by vessel, and this is
structural, independent of the exact values of $\sphase$ and the wall thickness
$h$. It is the strongest statement the present runs make, and it sets the agenda
for the per-vessel calibration of the companion study.

The model settles what limits the pump. In the range we explore the load,
not the drive, sets the ceiling. The transmural pressure and the passive wall
absorb most of what the muscle can raise, and the length-tension window pushes the
same way because every organ sits on its falling side. The drive still moves the
operating point, since pressure and outflow both rise with $\sphase$, but a
stronger muscle buys far less pressure than a lumped estimate predicts, and the
useful levers are the ones that lower the load.

The model resolves a standing ambiguity in the one pressure the clinic can
measure. A pressure read at the lymphovenous junction is not a clean contraction
signal, because the ostial valve stamps the respiratory rhythm onto it. This is
why the $\lesssim4$\,mmHg increase of Kelly \emph{et~al.}\ \cite{Kelly2022} admits
two readings, a respiration-led swing at the junction and a contraction-led swing
in the duct proper, and the model reproduces both at sizes consistent with that
measurement.

These results make the model a forward solver for lymphatic imaging. Dynamic
contrast-enhanced CT lymphangiography quantifies thoracic-duct flow
\cite{Molloi2023}, heavily $T_2$-weighted MR resolves the trunk
and its abdominal confluence \cite{Erden2005}, ultrasound tracks duct diameter and
motion under load \cite{Shinaoka2025}, and near-infrared fluorescence follows
propulsion in peripheral vessels \cite{groenlund2017validation,rasmussen2015near}.
All of these report transport or geometry and none reports vessel-level mechanics.
The model supplies the forward half of that inversion, predicting what a given
wall and valve parameter set produces at the sites these methods see, and the
contrast-tracer readout of the companion study closes it.

The load framing organizes several clinical observations. Lymphatic activity
falls during positive-pressure ventilation and general anesthesia, a proposed
cause of postoperative edema \cite{thorup2025reduced,Kelly2022}. Raising the
central venous pressure in swine lifts the junction pressure from $8.2$ to
$14.6$\,mmHg and remodels the duct \cite{Lu2022}, and the lymphatic reserve is
depressed in patients with a Fontan circulation \cite{Mohanakumar2021}. None of
these acts on the muscle and all act on the outlet, the side that sets the
amplitude here, and the model carries that sensitivity directly.

\subsection{Cerebrospinal-fluid drainage through the cervical route}
\label{sec:disc-cns}

The reach extends to the central nervous system, which drains into the same
conduit this model resolves. Cerebrospinal fluid exits the skull and drains
through a nasopharyngeal lymphatic plexus, subsequently flowing into deep cervical
lymphatics characterized by semilunar valves, long lymphangions, and smooth-muscle
coverage whose transport sets the outflow \cite{Yoon2024}. That segment is the
same object as every vessel
modelled here, a valved chain of contractile lymphangions, set apart only by where
it sits, at the head of the return path.

An injury model makes the two governing variables explicit. After traumatic
brain injury in mice an adrenergic surge lowers cervical lymphatic contractility
and raises the central venous pressure together, and pan-adrenergic blockade
normalizes the venous pressure, partly restores glymphatic and cervical lymphatic
flow, and reduces the edema \cite{Hussain2023}. Those two quantities are exactly
the drive and the load of this model, entering as the phasic-tension scale with
the wall timescale and as the venous reference of the ostial gate.

That shared boundary is a deliverable to those models. A cervical or
glymphatic model has to close its domain at the neck, so it prescribes a constant
pressure at the distal end of what it resolves. Kim and Tithof prescribe a murine
$6$\,mmHg there, swept to $10.44$\,mmHg in their sensitivity analysis
\cite{KimTithof2024}. This model computes that pressure, through the cervical
vessels and along the path that follows them, gated by the breath and modulated
by contraction. Their chain is murine and this
network is human. The murine value is the larger of the two while
$\alpha_p^{\mathrm{mouse}}$ scales murine pressures up to human ones, so neither
using the values as they stand nor rescaling them puts the two on one axis. A
venous blood pressure prescribed at a truncated boundary is also a different
object from a lymphatic luminal pressure computed inside the network.

At
the venous reference used throughout this paper, $P_{\mathrm{subclav}}=400$\,Pa,
the deep cervical node (at $v_{990}$, see figure~5 in \cite{Savinkov2020}) sits at $8.6$, $9.4$, $11.1$ and $13.8$\,mmHg for
$\sphase=5445$, $10000$, $15000$ and $30000$\,Pa, with a contraction swing of
$1.5$ to $5.6$\,mmHg. The superficial cervical median runs from $19.4$ to
$39.3$\,mmHg over the same range. Raising $P_{\mathrm{subclav}}$ to $2000$\,Pa
lifts the deep cervical node to between $21.1$ and $23.1$\,mmHg. The
sweep runs that high because traumatic brain injury raises the central venous
pressure above its resting value \cite{Hussain2023}, so the upper end of the range
is an injured state and not an arbitrary extrapolation.

\begin{figure}[t]
  \centering
  \includegraphics[width=\columnwidth]{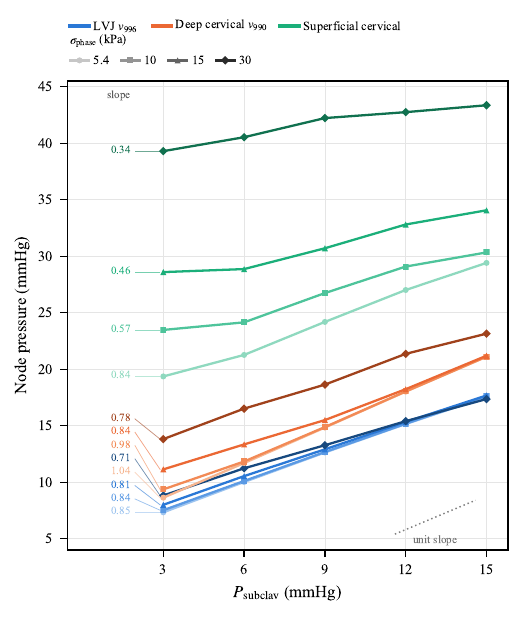}
  \caption{The venous boundary reaching the cervical drainage nodes. Every node is drawn against
  $P_{\mathrm{subclav}}$ with one line per $\sphase$. Color-function is the node and
  lightness for different $\sphase$. The three nodes are the lympho-venous
  junction $v_{996}$, the deep cervical node $v_{990}$ and the superficial
  cervical median. The number beside each line is that line's endpoint ratio over
  $P_{\mathrm{subclav}}=3$ to $15$\,mmHg, the fraction of a venous change that
  survives to the node, and the dotted segment shows what a ratio of one looks
  like on these axes. } 
  \label{fig:cns}
\end{figure}

The venous boundary reaches the neck undiminished when the muscle
is weak, and at the weakest drive it arrives amplified (Fig.~\ref{fig:cns}). Over the swept range the deep cervical pressure (the slope in Fig.~\ref{fig:cns}) rises by $1.04$, $0.98$,
$0.84$ and $0.78$ of the rise in $P_{\mathrm{subclav}}$ at
the vary $\sphase$ as given in last paragraph, and the superficial median by
$0.84$, $0.57$, $0.46$ and $0.34$. A ratio slightly above one is possible because
the outlet is a gate and not a fixed resistance, so raising $P_{\mathrm{subclav}}$
also throttles it and the net drainage falls by $23.5$\% across this sweep. The
junction itself then transmits only $0.85$, while the segment between the deep
cervical node and the junction takes up a larger drop, $1.3$ to $3.6$\,mmHg, and
that excess is what carries $v_{990}$ past unity.

Raising the venous pressure lifts every cervical node, but not with the same efficiency, and the efficiency is set by how hard the vessels contract. Write the node pressure as the venous pressure plus the head the contraction makes locally. The slope in Fig.~\ref{fig:cns} is then exactly one plus the change in that local head per unit change in the vein. When the contraction is weak the head is small and barely moves with the boundary, so nearly all of a venous change arrives. When the contraction is strong the head is large and raising the vein cuts into it, and the two changes partly cancel. On the other hand, the adrenergic surge of Ref.~\cite{Hussain2023} raises
$P_{\mathrm{subclav}}$ and lowers $\sphase$ at the same time, so both of its effects push the cervical pressure the same way.

At $\sphase=5445$,Pa the superficial median sits $16.4$,mmHg above the vein and loses $2.0$,mmHg of that head across the sweep, which leaves a slope of $0.84$. At $30000$,Pa it sits $36.3$,mmHg above the vein and loses $7.9$,mmHg, and the slope is $0.34$. The deep cervical node sits $5.6$ and $10.8$,mmHg above the vein at the two drives, and at the weaker one it gains $0.5$,mmHg of head instead of losing it, which is how its slope reaches $1.04$.

Kim and Tithof give their
reason for a murine model plainly, that anatomical data and flow measurements
for humans were not available to them. This model is built on what they record
as missing. Its graph is human anatomy, its passive wall law is fitted to
isolated human pelvic collecting lymphatic vessels \cite{Athanasiou2017}, its
venous reference is a human supine-resting central venous pressure, and its
calibre and outflow targets are human. What stays animal is the pumping, the
phasic tension, the frequency law, the duty ratio and the valve thresholds, all
rescaled from mouse and rat vessels, together with the ostial outlet resistance,
which is anchored to a swine thoracic duct \cite{Lu2022}. The human vessel data
that exists is passive, measured with the muscle tone suppressed, and that is
why the contractile side is animal in both models and why $\sphase$ is swept
here.

\begin{figure}[t]
    \centering
    \includegraphics[width=\columnwidth]{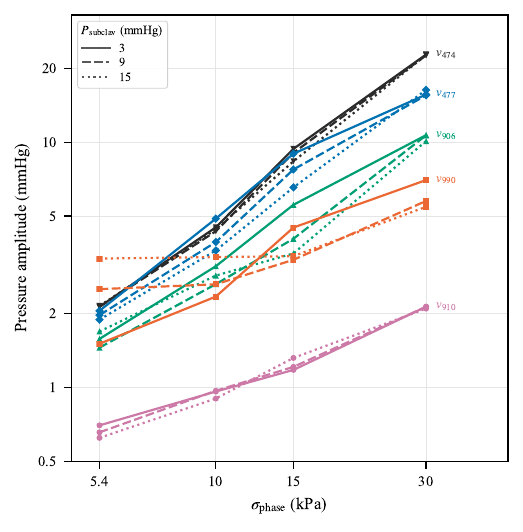}
    \caption{Pressure amplitude at the cervical outflow vertices. Colour is the node and line style is
    $P_{\mathrm{subclav}}$. The five nodes are the deep cervical node $v_{990}$ and
    the four head and neck collectors $v_{906}$, $v_{910}$, $v_{477}$ and $v_{474}$
    that the companion paper names. Both axes are logarithmic because the amplitude
    grows as a power of the drive and the five nodes do not share the exponent.
    $v_{990}$ keeps the orange it carries in Fig.~\ref{fig:cns}.}
    \label{fig:cnsamp}
\end{figure}

Raising $\sphase$ raises the strength of the pump, so it should raise the pressure oscillation the pump drives at every vessel as shown in Fig.~\ref{fig:cnsamp}. The amplitude at each node roughly follows a power law in $\sphase$, and the exponent depends on where the node sits. With varying $\sphase$, the amplitude at $v_{474}$ grows by a factor of $10.7$ with an exponent of $1.41$, and at $v_{910}$ by $3.0$ with an exponent of $0.65$. Changing $P_{\mathrm{subclav}}$ across its whole $3$ to $15$,mmHg range moves the four collector amplitudes  by at most $37\%$, against the factor of $3.0$ to $10.7$ that $\sphase$ moves them. On the other hand, the deep cervical $v_{990}$ sits closest to the vein, thus increase vein pressure effects it more. Its mean pressure stands only $5.3$,mmHg above $P_{\mathrm{subclav}}$ at the weakest drive, against $8.4$ to $22.9$,mmHg at the four collectors. Its amplitude rises by $197\%$ at $P_{\mathrm{subclav}}=3$,mmHg, by $32\%$ at $9$,mmHg and by $2\%$ at $15$,mmHg, which are the solid, dashed and dotted orange lines. 

\subsection{Limitations}
\label{sec:disc-limitations}

The main limitation is one of data. The parameters that
set the active tension, the passive wall, and the valve behaviour are not yet
measured across the human body, so the model cannot report the true parameter
distribution of the whole human lymphatic system, and even the vessel geometry
rests on a graph that more anatomical data would refine. For this reason we study
a single healthy configuration here and leave the diseased states, where these
parameters shift, to future work once the measurements exist. We read this
healthy, whole-body model as a foundation to build on, the start of a longer
effort.

A few model choices bound the results more narrowly. The phasic-tension
scale and the valve sharpness are held globally constant, so the runs show the
response of a uniform drive and not the per-vessel calibration of a separate
study. The runs reach a quasi-steady operating point at the lymphovenous junction
over $200$\,s and not a whole-network steady state. The graph does not resolve the
smallest vessels, whose flow direction produces the long tails of
Fig.~\ref{fig:amp}. None of the clinical statements above is tested against
patient data, and the contrast-tracer readout that would carry that test is
deferred to a companion paper.

The two cervical numbers do not carry the same weight. The graph labels
one vertex as deep cervical, $v_{990}$, so every deep cervical value here is a
single node with a single outflow edge, while the superficial median is taken
over $140$ labelled vertices. The part of the network closest to the central
nervous system is also the least resolved part of the graph.

The model stops short of the intracranial source, by construction. It carries no meningeal or perivascular compartment and no cerebrospinal-fluid production term, and it resolves the whole deep cervical chain with a single node. The murine model of Kim and Tithof resolves that same segment as a chain of five lymphangions \cite{KimTithof2024}, and glymphatic network models esolve the compartments upstream of it \cite{Tithof2022,Bohr2022}. What this model adds is everything downstream, a whole-network load and a venous outlet that is computed, and it pays for that with the cervical detail.

The most consequential limitation of the cervical results is the choice of
boundary condition. Peripheral lymph formation enters as a prescribed inflow at
every afferent leaf, so the network is driven at fixed flow and not at fixed
pressure, and raising $P_{\mathrm{subclav}}$ then lifts the deep cervical pressure
almost in proportion, which is the near-unit slope of the deep cervical
lines in Fig.~\ref{fig:cns}. That near-linear response is a property of the boundary
condition and may not be of the anatomy, and every transmission ratio in
Fig.~\ref{fig:cns} is conditional on it. A patient does not have to behave this way, for example, in oedema the
interstitial compartment acts as a reservoir that holds a pressure instead of
delivering a fixed flow, so a prescribed-pressure inflow is likely the better
boundary there, and the ratios would change with it. Choosing the boundary
condition to match a given clinical state is left to future work.

\section{Conclusion}
\label{sec:conclusion}

We present the first model to place active lymphangion pumping on an
anatomically derived human whole-body lymphatic graph \cite{Savinkov2020}. It
advances the passive network of the companion study \cite{Chang2026Static} into
the contraction-driven regime, coupling intrinsic lymphangion contraction, finite
wall relaxation, dynamic secondary valves, the swept-volume source of each
shortening segment, and a respiration-gated venous outlet, and advancing the
coupled hydraulic system in time. This gives lymphatic imaging a forward model for reading
network-level measurements in terms of vessel-level mechanics, and gives clinical
research a platform for asking where a failing pump first loses its reserve. For a
system that resists direct measurement almost everywhere in the body, it is a step
toward a quantitative, whole-body account of human lymphatic function.

The impact is that the contraction of the human thoracic duct is now
simulated as it happens in the body, and that this gives the model the reach to
predict disease. Cerebral oedema is one such case, where the lymph behind the
swelling has to drain through the cervical route into the thoracic duct, and this
solver returns that drainage as a pressure and flow field.

\section*{List of abbreviations}
CC: cisterna chyli;
CNS: central nervous system;
CSF: cerebrospinal fluid;
GI: gastrointestinal;
ICG: indocyanine green;
LN: lymph node;
LVJ: lymphovenous junction;
NIRF: near-infrared fluorescence;
RCR: resistance--compliance--resistance (three-element Windkessel);
TD: thoracic duct.

\section*{Declarations}

\subsection*{Ethics approval and consent to participate}
Not applicable.

\subsection*{Consent for publication}
Not applicable.

\subsection*{Availability of data and materials}
Data and materials are available from the corresponding author on
reasonable request.

\subsection*{Competing interests}
The authors declare no competing interests.

\subsection*{Funding}

\subsection*{Authors' contributions}
Conceptualization: CFC; Methodology: CFC; Formal analysis and investigation:
CFC, BA, ES; Writing - original draft preparation: CFC; Writing - review and editing:
CFC, BA, ES, SM; Funding acquisition: SM; Resources: SM; Supervision: SM.

\subsection*{Acknowledgements}
Not applicable.

\begin{table*}[t]
\centering
\small
\caption{
Macro-scale collecting-lymphatic pumping-frequency observations by region/preparation and reported pressure metric.
The listed frequencies denote experimentally reported event/contractile frequencies in the corresponding preparation.
The pressure metrics are not a single uniform observable across rows:
they include estimated occlusion/pumping pressure (human peripheral NIRF studies),
invasive luminal pressure and contraction-induced pressure rise (thoracic duct),
and ex vivo equivalent transmural pressure at maximal active tension (wire-myograph studies).
NR: not reported.
}
\label{tab:lymph_pumping_frequency_updated}
\resizebox{\textwidth}{!}{%
\begin{tabular}{cccccc}
\hline
\makecell[c]{Region / preparation}
&
\makecell[c]{frequency\\ (min$^{-1}$)}
&
\makecell[c]{pressure metric}
&
\makecell[c]{evidence type}
&
reference
&
note \\
\hline

\makecell[c]{Upper extremity superficial\\ collectors}
&
$0.9 \pm 0.4$
&
\makecell[c]{estimated occlusion / pumping\\ pressure: $59 \pm 12$ mmHg\\
whole-limb functional index;\\
not direct luminal or transmural pressure}
&
\makecell[c]{human direct\\ (in vivo)}
&
\cite{Kelly2020}
&
\makecell[c]{Healthy males; superficial\\ collectors imaged by NIRF.} \\
\hline

\makecell[c]{Lower extremity superficial\\ collectors}
&
$0.4 \pm 0.3$
&
\makecell[c]{estimated occlusion / pumping\\ pressure: $54.7 \pm 9.4$ mmHg \\
whole-limb functional index;\\
not direct luminal or transmural pressure}
&
\makecell[c]{human direct\\ (in vivo)}
&
\cite{Mohanakumar2020}
&
\makecell[c]{Healthy postmenopausal women;\\ placebo baseline.} \\
\hline

\makecell[c]{Lower extremity superficial\\ collectors}
&
$0.3 \pm 0.2$
&
\makecell[c]{estimated occlusion / pumping\\ pressure: $57.2 \pm 8.4$ mmHg\\
whole-limb functional index;\\
not direct luminal or transmural pressure}
&
\makecell[c]{human direct\\ (in vivo)}
&
\cite{Mohanakumar2021}
&
\makecell[c]{Healthy controls in the Fontan\\ comparison study.} \\
\hline

\makecell[c]{Lower extremity superficial\\ collectors}
&
$0.5 \pm 0.2$
&
\makecell[c]{standing vs supine:\\ $+9$ mmHg\\
whole-limb functional index;\\
not direct luminal or transmural pressure}
&
\makecell[c]{human direct\\ (in vivo)}
&
\cite{HolmWeber2022}
&
\makecell[c]{Posture study; reports pressure\\ change with standing, not a\\ single absolute baseline value.} \\
\hline

\makecell[c]{Thoracic duct\\ (central duct)}
&
$5.0 \pm 1.0$
&
\makecell[c]{luminal pressure: $16 \pm 5$ mmHg;\\ contraction-induced rise:\\ $\Delta P_{\mathrm{contr}}=4 \pm 4$ mmHg}
&
\makecell[c]{human direct\\ (invasive)}
&
\cite{Kelly2022}
&
\makecell[c]{Central thoracic duct; severe\\ congenital-heart-disease cohort;\\ not a peripheral superficial\\ collector.} \\
\hline

\makecell[c]{Human intestinal mesenteric\\ lymphatic vessels (MLVs)}
&
$3.4 \pm 1.7$
&
\makecell[c]{equivalent transmural pressure at\\ maximal active tension:\\ $22$ mmHg}
&
\makecell[c]{human direct\\ (ex vivo)}
&
\cite{Telinius2017}
&
\makecell[c]{Isolated human intestinal MLVs\\ mounted in wire myograph;\\ pressure value is an equivalent\\ transmural quantity, not in vivo\\ pumping pressure.} \\
\hline

\makecell[c]{Porcine mesenteric\\ lymphatics}
&
$1.7 \pm 0.2$
&
NR
&
\makecell[c]{large-mammal surrogate\\ (ex vivo)}
&
\cite{Ferguson1994}
&
\makecell[c]{Ring preparation with\\ spontaneous contractions.} \\
\hline

\makecell[c]{Porcine tracheobronchial\\ lymphatics}
&
$0.6 \pm 0.2$
&
NR
&
\makecell[c]{large-mammal surrogate\\ (ex vivo)}
&
\cite{Ferguson1994}
&
\makecell[c]{Ring preparation; only a subset\\ of rings showed spontaneous\\ activity.} \\
\hline

\makecell[c]{Head / neck / Superficial trunk\\superficial collectors\\ (working prior)}
&
$\sim 0.65 \pm 0.21$
&
\textemdash{}
&
\makecell[c]{weak inference\\ only}
&
\cite{Kelly2020,Mohanakumar2020,Mohanakumar2021,HolmWeber2022}
&
\makecell[c]{No direct region-specific\\ measurement was identified.\\
A working prior was therefore\\
constructed from human\\
peripheral superficial collecting-\\
vessel data by averaging the\\
upper-extremity value\\
($0.9 \pm 0.4$ min$^{-1}$) and the\\
pooled lower-extremity value\\
($0.40 \pm 0.12$ min$^{-1}$), yielding\\
$\sim 0.65 \pm 0.21$ min$^{-1}$.} \\
\hline

\end{tabular}%
}
\end{table*}

\bibliographystyle{vancouver}
\bibliography{References}

\end{document}